\documentclass{aa}
\usepackage{natbib}
\bibpunct{(}{)}{;}{a}{}{,} 
\pdfoutput=1
\usepackage{txfonts}
\usepackage{graphicx}
\newcommand{\bm}[1]{\mbox{\boldmath$#1$\unboldmath}}

\begin{document}
\title{Bright fibrils in \ion{Ca}{II} K}
\author{A.~Pietarila\inst{1} \and J.~Hirzberger\inst{1} \and V. Zakharov\inst{1} \and S.K.~Solanki\inst{1,2}}
\institute{Max-Planck-Institut fur Sonnensystemforschung, 37191 Katlenburg-Lindau, Germany \and School of Space Research, Kyung Hee University, Yongin, Gyeonggi, 446-701, Korea} 
\date{Received <date> / Accepted <date>}

\abstract {Except for the \ion{Ca}{II} resonance lines, fibrils are ubiquitously present in most high-resolution observations of chromospheric lines.}{We show that fibrils are also a prevailing feature in \ion{Ca}{II} K, provided the spatial-resolution is sufficiently high.}{We present high spatial resolution observations of an active region in the \ion{Ca}{II} K line from the Swedish Solar Telescope. Through a comparison between photospheric intensity and magnetic field data, we study the connection between bright chromospheric fibrils and photospheric structures. Additionally, using Fourier analysis we study how the fibrils are linked to the observed dynamics.} {We find that very narrow, bright fibrils are a prevailing feature over large portions of the observed field. We also find a clear connection between the fibril footpoints and photospheric magnetic features. We show that the fibrils play two distinct roles in the observed dynamics: depending on their location they can act as a canopy suppressing oscillations or they can channel low-frequency oscillations into the chromosphere.}{The \ion{Ca}{II} K fibrils share many characteristics with fibrils observed in other chromospheric lines, but some features, such as the very small widths, are unique to these observations.}


\maketitle

\titlerunning{<Fibrils in \ion{Ca}{II} K>}
\authorrunning{<Pietarila et~al.>}
\section{Introduction}

The chromosphere is very inhomogeneous, both in space and
time. When viewed in strong spectral lines, such as H$\alpha$, the presence
of fibril-structures is evident everywhere. They are usually called spicules when located on the solar limb, mottles in the
quiet Sun, and fibrils in active regions. Regardless of the name, these
features are closely related to each other. Spicules carry a mass flux
that is 100 times that of the solar wind into the lower corona
\citep{Withbroe1983}, making them an important factor in the mass
balance of the solar atmosphere. The dynamic properties of many
fibrils and spicules are consistent with the driver being photospheric
oscillations that are leaking into the chromosphere along inclined magnetic
field lines and forming shocks that in turn drive chromospheric jets
\citep{Hansteen+others2006}. Hinode SOT observations have revealed
a new subset of spicules which are even more dynamic, and for which a
reconnection-related formation mechanism is a likely explanation
\citep{DePontieu+others2007}. In addition, there are also heavily
inclined, low-lying fibrils that are more static and do not show
jet-like behavior \citep{DePontieu+others2007b}. In all fibril types the
observed structures (e.g., length) and dynamics (e.g., periodicity,
lifetime) are clearly tied to the local magnetic field
configuration.

 Besides their ubiquitous presence in H$\alpha$, fibrils are seen in most chromospheric lines, e.g.,
\ion{Ca}{II} 8542 \AA\ (\citealt{Vecchio+others2007}, \citealt{Cauzzi+others2008}) and Ly-$\alpha$ 1216
\AA\ \citep{Patsourakos+others2007}. However, prominent on-disk fibrils in the \ion{Ca}{II} H and K lines are seldom observed. Bright and dark fibrils have been decribed by \cite{Zirin1974} in narrowband (0.3 \AA) \ion{Ca}{II} K filtergrams of the quiet Sun and active regions (see also \citealt{Marsh1976}), but more often observations of the \ion{Ca}{II} K and H lines typically show long thin emission features, ``straws'' \citep{Rutten2007} only close to the limb. On the disk observations usually show a structure
of diffuse brightenings in magnetic regions and an even more complex pattern in quieter regions. In this paper we show that
the fibril structures are ubiquitous in the \ion{Ca}{II} K line as
well, if observed at sufficiently high spatial resolution, and in a sufficiently narrow wavelength band to avoid contamination by photospheric radiation. The latter issue has been discussed in detail by \cite{Reardon+others2007}.

The paper is organized as follows: in the next section the data and
its reduction as well as analysis are discussed in some detail. Then
the results, bright \ion{Ca}{II} K fibril structures, widths and dynamics,
are presented. The results are discussed and put into context with
observations of other spectral lines in section 4. Finally, section 5
summarizes the main results.

\section{Observations, data reduction and analysis}

A region close to solar disk center (AR 10966, $\mu=0.99$) was observed at the Swedish Solar Telescope
(SST, \citealt{SST2003}) on Aug. 9, 2007 from 13:31 to 14:12 UT in
periodically excellent seeing conditions. The observations were made
using the adaptive optics (AO, \citealt{SSTAO}) simultaneously with a 1.5 \AA\ wide
\ion{Ca}{II} K (centered at line core) and 1 nm wide continuum interference filter (centered at
395.37 nm). The analyzed
\ion{Ca}{II} K and continuum image sequences are composed of 165 frames (10 ms exposure time) each with a
cadence of 15 s. The SOUP tunable
birefringent filter \citep{Title+Rosenberg1981} was used to record the
full Stokes profile of the
photospheric Fe I 6302 \AA\ line at six wavelength positions ($-250$,
$-150$, $-75$, $0$, $75$ and $150$ m\AA\ from line core) with a cadence of 123 s for the full set of 4 polarization states and for the six positions. The SOUP images
were reconstructed using speckle reconstruction (e.g,
\citealt{Weigelt1977}, \citealt{deBoer1996}). Multi-object
multi-frame blind deconvolution (MOMFBD, \citealt{MOMFBD2005}) and
speckle reconstruction were independently applied to the interference
filter data to obtain images with very high spatial resolution. The Fried parameters and hence the applied speckle transfer
functions were estimated in dependence of the distance of the
individual subfields from the lock point of the AO (see \citealt{Puschmann+Sailer2006}).

The two different image reconstruction methods yield very similar
results for the continuum images, but there is a discrepancy between
the two in the \ion{Ca}{II} K images: the MOMFBD results in images in
which sharp fibrils lie on a more uniform hazy component, whereas the
speckle reconstructed images do not show this uniform feature. In the
following, only properties found in images from both reconstruction
methods are discussed. Furthermore, we focus exclusively on the bright fibril structures and do not address the dark structures in the present study.

The pixel size of the \ion{Ca}{II} K and continuum data is 0.033
$\times$ 0.033 arcseconds and for the SOUP data 0.065 $\times$ 0.065
arcseconds. The field of view (FOV) is roughly 60 $\times$ 60 arcseconds for
both the \ion{Ca}{II} K and continuum data. It is a few arcseconds smaller for the
SOUP data. In the best frames the spatial resolution of the
reconstructed \ion{Ca}{II} K and continuum images is diffraction
limited (the Rayleigh limit for Ca II K is 0.1 arcseconds). The
spatial resolution of the SOUP images is somewhat lower, approximately
0.2 arcseconds. 

For the SOUP data instrumental polarization effects of the laboratory
setup were measured with dedicated calibration optics and the telescope
polarization was determined using a model developed by \cite{Selbing2005}.
The thus deduced demodulation matrices were applied to the SOUP
data resulting in the full Stokes vectors at six spectral positions
for each pixel of the FOV. The Stokes profiles were inverted assuming a one-component-plus-straylight
Milne-Eddington atmosphere using
the inversion code, HeLiX (\citealt{Lagg+others2004}). The magnetic component of the obtained atmosphere includes the magnetic field vector,
${\bm B}=(\left|{\bm B}\right| ,\gamma ,\chi )$, where $\gamma$ is the
field inclination with respect to the line of sight and $\chi$ is the
azimuth angle of the magnetic field, as well as the magnetic filling
factor, $\alpha$, and the line-of-sight flow velocity, $v$. In the straylight component, having filling factor $(1-\alpha )$, the magnetic
field vector is assumed to be zero and all other parameters are
coupled to the values of the magnetic component. This is a particularly simple way of accounting for the fact that only part of the rather small resolution element is filled with fields and we do not know the local atmospheric structure, except that it is unlikely to be well represented by the quiet Sun. From the inversion
results we calculated maps of the Doppler velocity, $v$, and of the
longitudinal magnetic flux density,
$\left|{\bm B}\right|\cos\gamma\cdot\alpha$, which are the two most
robust quantities.     

\begin{figure*}
\begin{center}
\includegraphics[width=10cm]{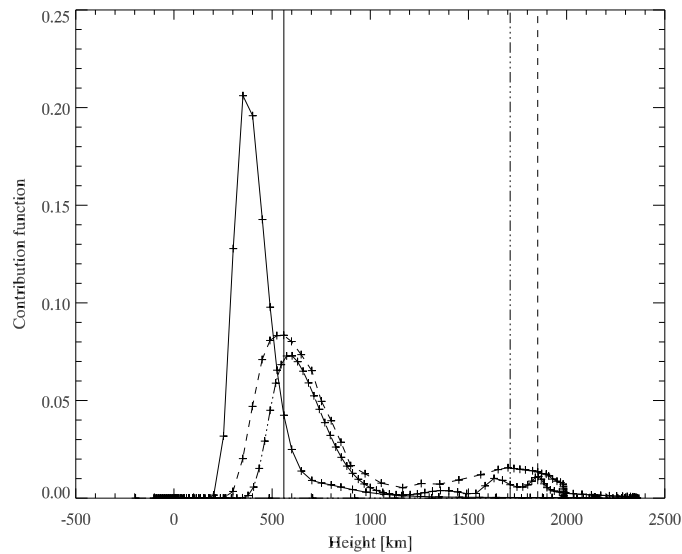}
\caption{Contribution functions for the \ion{Ca}{II} K filter used in
  the observation for the FALC model(quiet Sun \citealt{FAL1993} represented by the solid line), FALF model (network, \citealt{FAL1993}; in dashed line)
  and a Carlsson \& Stein atmospheric model snapshot from a 1-dimensional radiative hydrodynamics simulation of a shock wave propagating through
  the formation region of the \ion{Ca}{II} K line (\citealt{Carlsson+Stein1997};
   in dash-dotted line. The vertical dotted lines show the location from below which
  90 \% of the contribution is coming. Units of the y-axis are
  arbitrary. }\label{fig:0}
\end{center}
\end{figure*}

The \ion{Ca}{II} K filter contribution function in the quiet Sun (FALC
model of \citealt{FAL1993}) shows that most of the signal is coming
from the upper photosphere (Fig.~\ref{fig:0}). In fact, 90\% of the
emission in a FALC atmosphere is coming from below 550 km. The
contribution functions in Fig.~\ref{fig:0} were calculated using the
nLTE radiative transfer code of H. Uitenbroek
(\citealt{Uitenbroek1998}, http://www.nso.edu/staff/uitenbr/rh.html). The code uses the
Multi-level Accelerated Lambda Iteration (MALI) formalism of Rybicki
\& Hummer (\citealt{Rybicki+Hummer1991, Rybicki+Hummer1992})
and takes into account effects of partial frequency redistribution for
the \ion{Ca}{II} K line. Since no transmission curve was available for
the \ion{Ca}{II} K filter, the contribution function was calculated
for a 1.5 \AA\ wide rectangular bandpass centered at the line core. The
contribution function extends to the FALC chromosphere, but the
chromospheric contribution is at best minor, especially at heights
above 1000 km. In comparison, in regions such as the network and plage (described by FALF, i.e., model F of \citealt{FAL1993})
a much larger portion of the emission arises from the chromosphere,
see dashed line in Fig.~\ref{fig:0}, which corresponds to the filter
contribution function for the FALF model. Noticeable is the presence
of a second peak in the middle to upper chromosphere. Also an upward
propagating shock wave such as in the simulations of Carlsson \& Stein
\citep{Carlsson+Stein1997} produces a secondary chromospheric peak in the
contribution function (dash-dotted line in
Fig.~\ref{fig:0}). Consequently the data mainly sample the upper
photosphere and temperature minimum region in quiet areas, but become
sensitive to chromospheric features as soon as the temperature there
is high, i.e., above photospheric magnetic features and above a hot
canopy that may surround them. Note that Fig.~\ref{fig:0} only gives a qualitative picture of the heights from which the emission in the \ion{Ca}{II} K filter is coming, since none of the employed atmospheres can be deemed to be entirely realistic. The contribution functions presented here are comparable to the response functions presented by \cite{Carlsson+others2007} for the Hinode SOT Ca II H filter. That filter is somewhat wider than the filter used in this study and, consequently, the chromospheric contributions are slightly smaller.

\begin{figure*}
\begin{center}
\includegraphics[width=16cm]{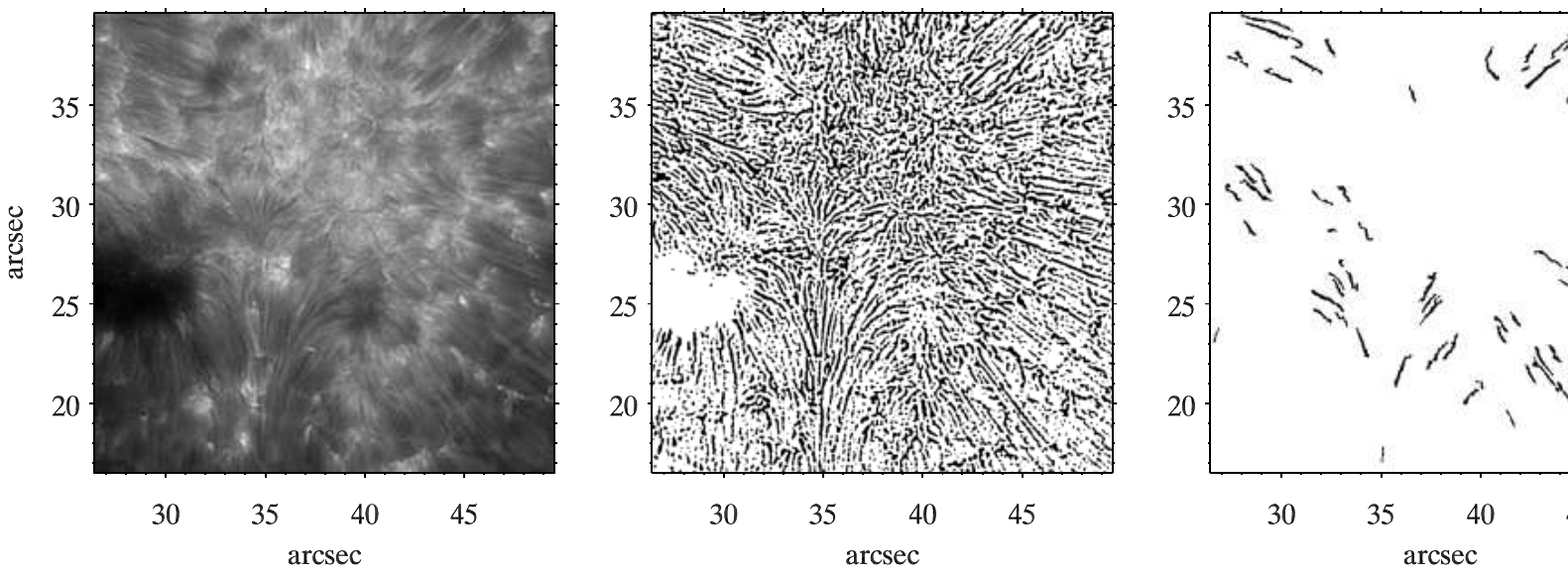}
\caption{Automatic detection of fibrils. See text for details. First
  image: original intensity image (linear color scale). Second: first binary image. Third:
  identified fibril segments.  }\label{fig:9}
\end{center}
\end{figure*}

To isolate the bright fibrils in the intensity images the
following procedure was applied to 12 of the highest quality \ion{Ca}{II} K
images. First an intensity threshold was applied (5000-17000 counts)
and the resulting image was filtered with an unsharp mask (radius 5
pixels). Next a second threshold was applied (210-1990 counts)
and the resulting image was converted from float to binary. The standard IDL procedure
${\rm label\_region}$ was run on the binary image. The procedure
labels each region (a non-zero valued pixel and all the adjacent neighboring non-zero pixels) in the image. Each of the labeled regions were
studied individually and ones fulfilling certain criteria were
kept. The criteria were: the region must consist of at least 60
pixels. The longest distance between two points must be at least 25
pixels. These two criteria are used to remove features that are too small
in size. Additional criteria include the ratio of the circumference of
the region to the longest distance between two points in the region
must be less than 2.8. In addition, the edges of each region were identified and
the slope between adjacent pixel pairs computed. Only regions, where
the ratio of the mean of the absolute value of the slope and the
standard deviation is larger than $0.9$, are kept. These last criteria are employed
to remove features that are not fairly straight and elongated, e.g.,
granule edges. The regions in the resulting binary image are again
identified with the ${\rm label\_region}$ procedure and regions with
more than 55 pixels are identified as segments of fibrils. The method outlines the fibrils and enables computation of their widths. The masking
and criteria were found through trial and error. In Fig.~\ref{fig:9}
are shown a portion of an original image, the first binary image,
and the identified segments. The procedure creates only very few false
identifications, but it misses many fibrils. Nonetheless, the 12 processed
images resulted in nearly 6000 bright fibril segments and the properties of
the segments are consistent in all the images. This confirms that
the detection method is not sensitive to small changes in spatial
resolution provided that the spatial resolution of the original images
is sufficiently high.

A second way, entirely independent of the above described algorithm, to define fibril widths is to manually identify the edges
of a fibril segment from the intensity images and calculate the width from that (shortest
distance between the segment edges). This was done for four regions of a single snapshot,
and served mainly to compare with and test the results obtained with the
automated method. No measurements of full width at half maximum were made because the background intensity pattern also varies on very small scales and in many places the fibrils overlap each other.

To sketch out the large scale structure of the fibrils a \ion{Ca}{II}
K intensity image was divided into sections of $100 \times 100$ pixels
($3.3^{\prime \prime} \times 3.3^{\prime \prime}$). Then each section
was displayed and a dominant fibril direction was determined by
eye. If no fibrils were present, no direction was determined. This was
done for two intensity snapshots taken 36.5 minutes apart. The two
images were chosen since they have the best spatial resolution and are
located far away from each other in time.

To study the spatial properties of intensity oscillations standard
Fourier techniques were used. We applied the Fourier transform on intensity fluctuations normalized to the temporal mean. Without such a normalization the power maps would be dominated by intensity fluctuations of the highest intensity pixels. The resulting power spectra are divided
into three frequency regions: 0.4-1.6 mHz, 2.4-4 mHz and 5.2-8
mHz. The first frequency region encompasses the long time evolution
($>15$ min), the second is dominated by the photospheric 5 minute
oscillations (normally filtered out by the acoustic cutoff frequency in the chromosphere), and the third
one is (often considered mainly chromospheric) oscillations above the acoustic cutoff
frequency. Because varying seeing conditions cause peaks in the
power spectra it is not possible to compare the power in different
frequency ranges. However, since the effect of the seeing is roughly
the same for the continuum and \ion{Ca}{II} K intensities in all the
pixels, it is still possible to compare different spatial regions.

\section{Descriptive results}

\subsection{Large scale pattern}

\begin{figure*}
\includegraphics[width=11.5cm]{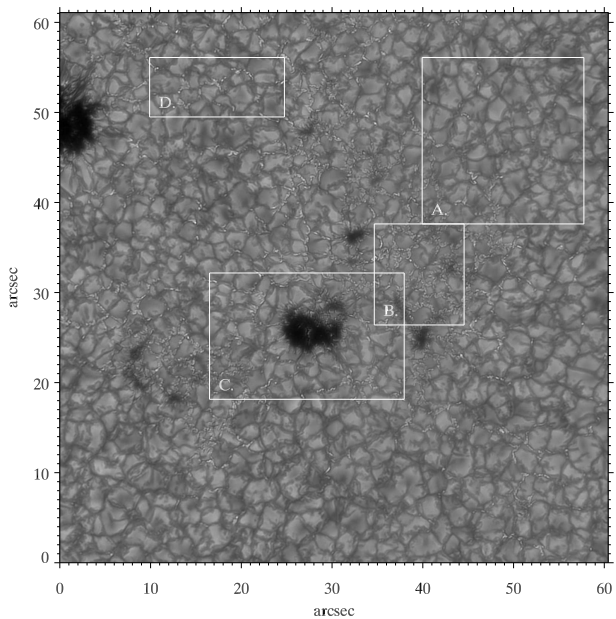}
\caption{Continuum intensity image obtained at 13:24:05-13:24:20 UT. The four boxes mark regions
  discussed later in the text. The color scale is linear.}\label{fig:1a}
\end{figure*}

\begin{figure*}[P]
\begin{center}
\includegraphics[width=13.7cm]{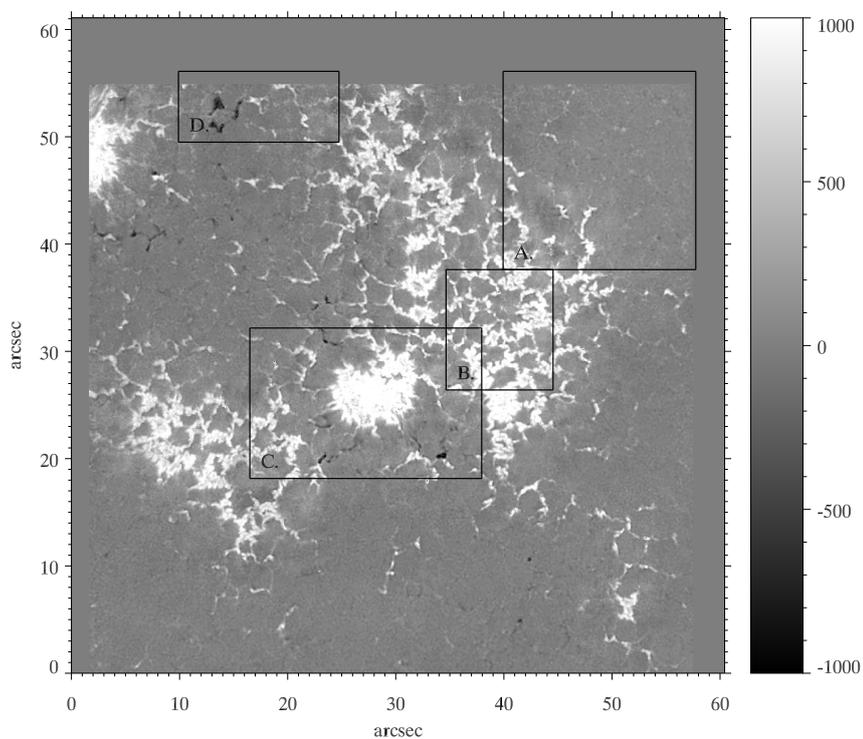}
\caption{As Fig.~\ref{fig:1a} but for the longitudinal magnetic flux
  (in G) in the photosphere. Note that this image, recorded at 13:25:12-13:27:19 UT, is not cotemporal with Fig.~\ref{fig:1a}.}\label{fig:1c}
\end{center}
\end{figure*}

\begin{figure*}[P]
\begin{center}
\includegraphics[width=13.7cm]{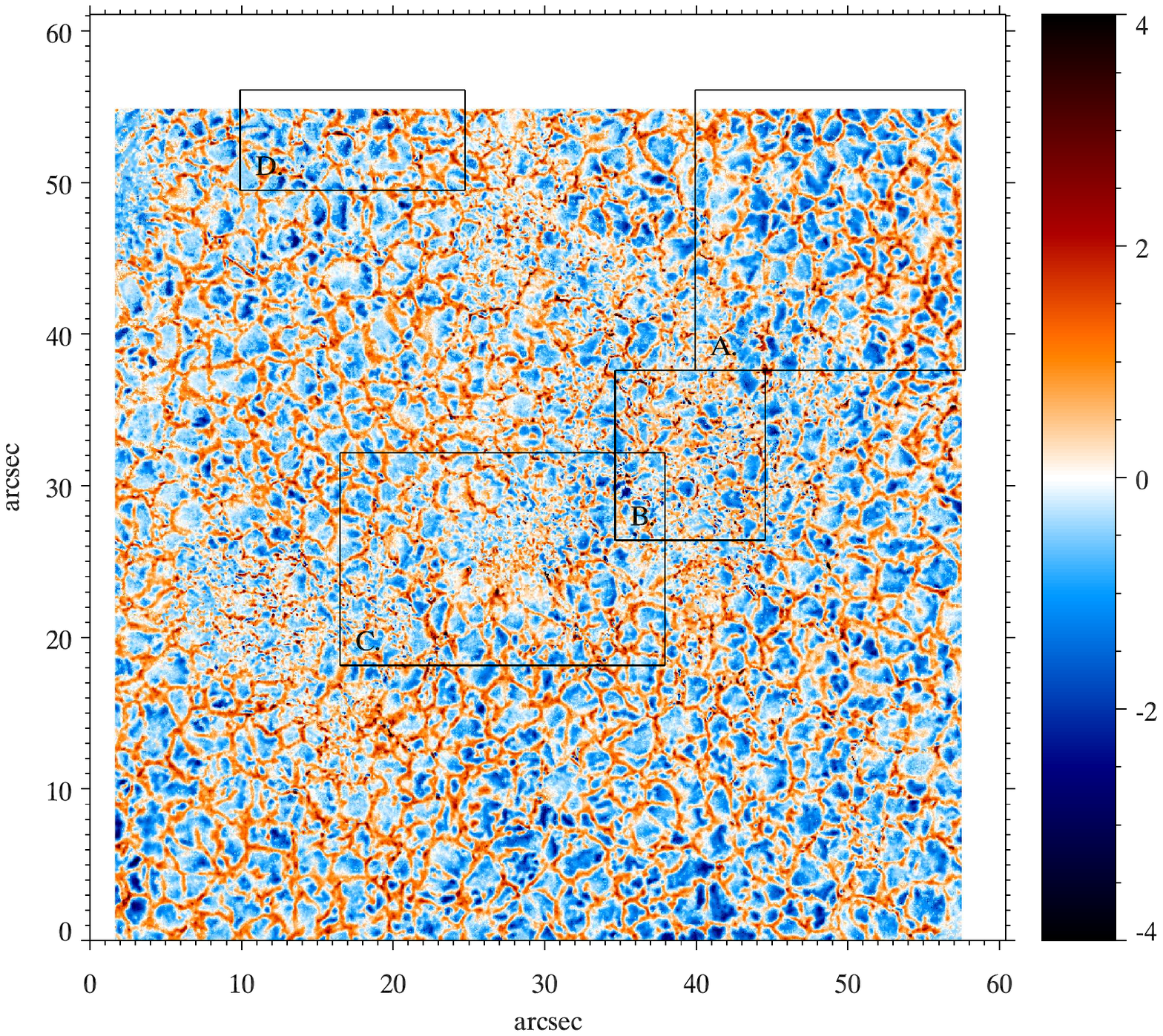}
\caption{As Fig.~\ref{fig:1a} but for the photospheric velocity (in km
  $\mathrm s^{-1}$).  Note that this image, recorded at 13:25:12-13:27:19 UT, is not cotemporal with Fig.~\ref{fig:1a}.}\label{fig:1d}
\end{center}
\end{figure*}

The observed region is a decaying active region, AR 10966, located
close to disk center. In continuum intensity the center of the FOV is dominated by a pore surrounded by plage
(Fig.~\ref{fig:1a}) where many of the intergranular lanes are filled with bright filigree. A second pore is located in the upper left corner
of the FOV. The darker regions in the plage are
micropores. Some isolated bright points are located in the quiet Sun
at the lower and right edges of the FOV. The agreement between the
photospheric intensity and the photospheric magnetic field obtained from a Milne-Eddington inversion
(Fig.~\ref{fig:1c}) is good: nearly all of the bright photospheric
emission can be identified as regions with significant
magnetic flux. Most of the FOV is unipolar though small patches of
opposite polarity are present in the upper left corner in between the
two pores and also below the central pore in the middle of the
FOV. The pores are nearly isolated from the plage: they are surrounded
by a layer of granulation with very little magnetic flux. Only the
upper right corner of the central pore is directly connected to the
plage. The granulation pattern is abnormal throughout most of the
plage region, being composed of particularly
small granules and intergranular lanes filled with filigree. Often, but not always, the red-shift relative to the rest of the FOV) covers more of the surface in the abnormal granulation than in the more quiet regions
(Fig.~\ref{fig:1d}). In addition, the downflow lanes in the abnormal granulation appear more patchy, with strong downflows restricted to individual point-like features. 


\begin{figure*}[P]
\begin{center}
\includegraphics[width=11.5cm]{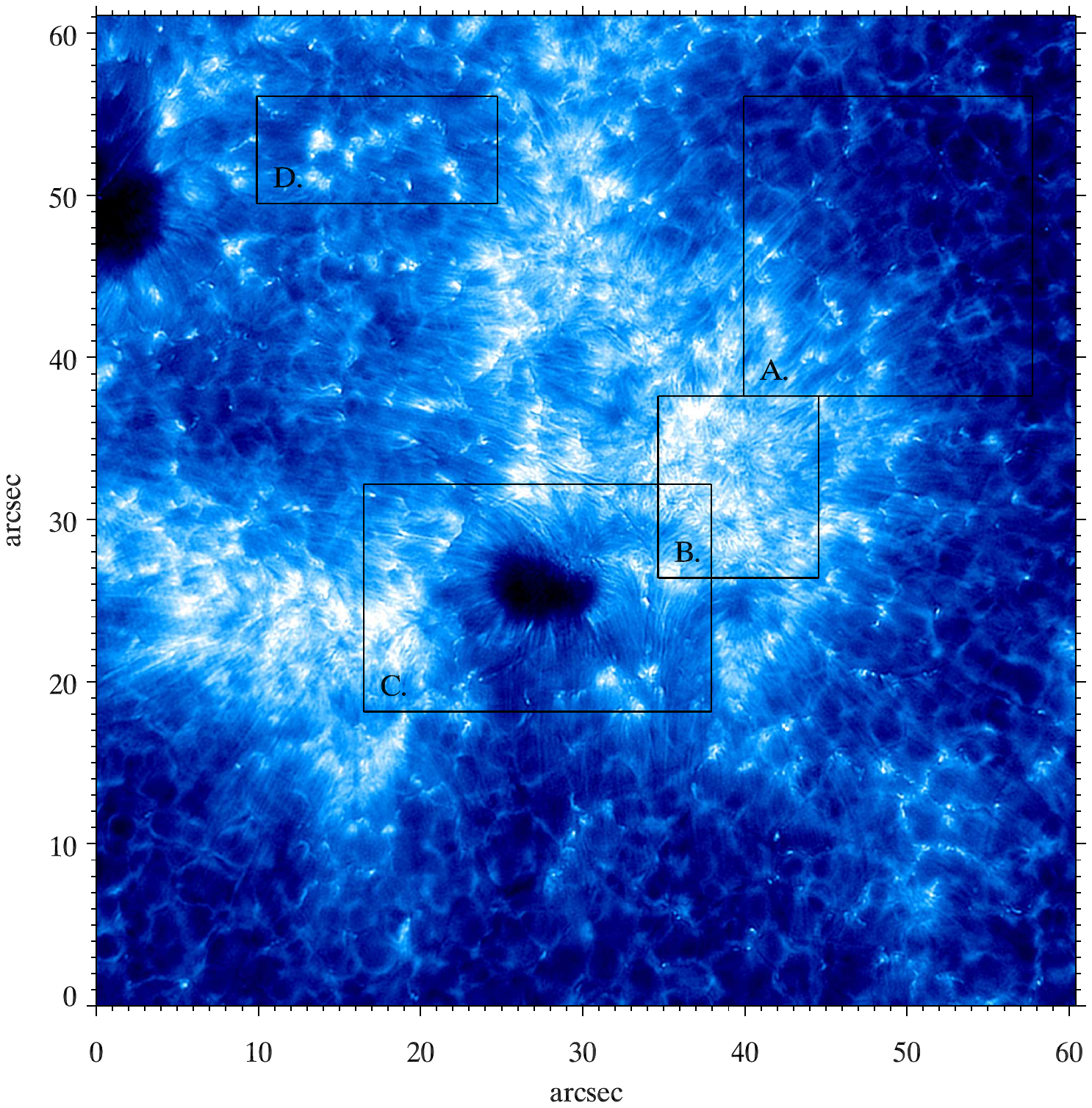}
\caption{As Fig.~\ref{fig:1a} but for \ion{Ca}{II} K obtained at 13:24:05-13:24:20 UT. Color scale is linear}\label{fig:1b}
\end{center}
\end{figure*}

The pores and plage are clearly visible in \ion{Ca}{II} K intensity
(Fig.~\ref{fig:1b}). The most distinct chromospheric feature is the bright
fibrils originating from magnetic regions. The end points of the
fibrils are cospatial with the photospheric bright points and magnetic
concentrations. In the plage the stalky, short fibrils form a thick,
carpet-like covering whereas at the plage edges the fibrils are
longer, more organized and often nearly parallel to one another. The
long fibrils extend over multiple granules and gradually fade out of
view over the quiet Sun. The patch of network visible in the lower
right corner of the FOV is also associated with fibrils which are much
shorter than the fibrils originating from the plage edge. Outside the
plage the dominant structure is reversed granulation \citep{Cheung+others2007, Janssen+Cauzzi2006}.

\begin{figure*}[P]
\begin{center}
\includegraphics[width=6cm]{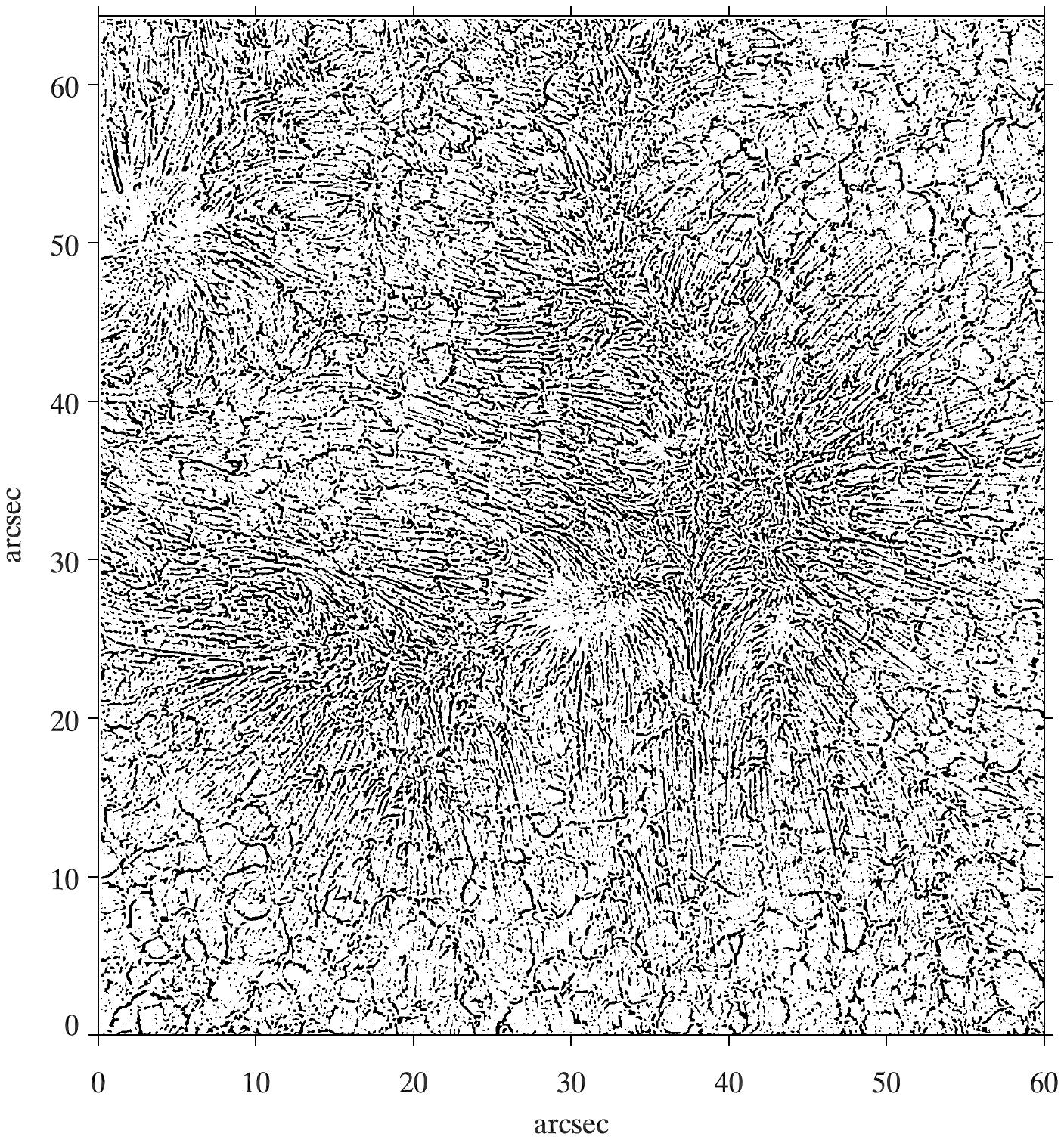}
\includegraphics[width=6cm]{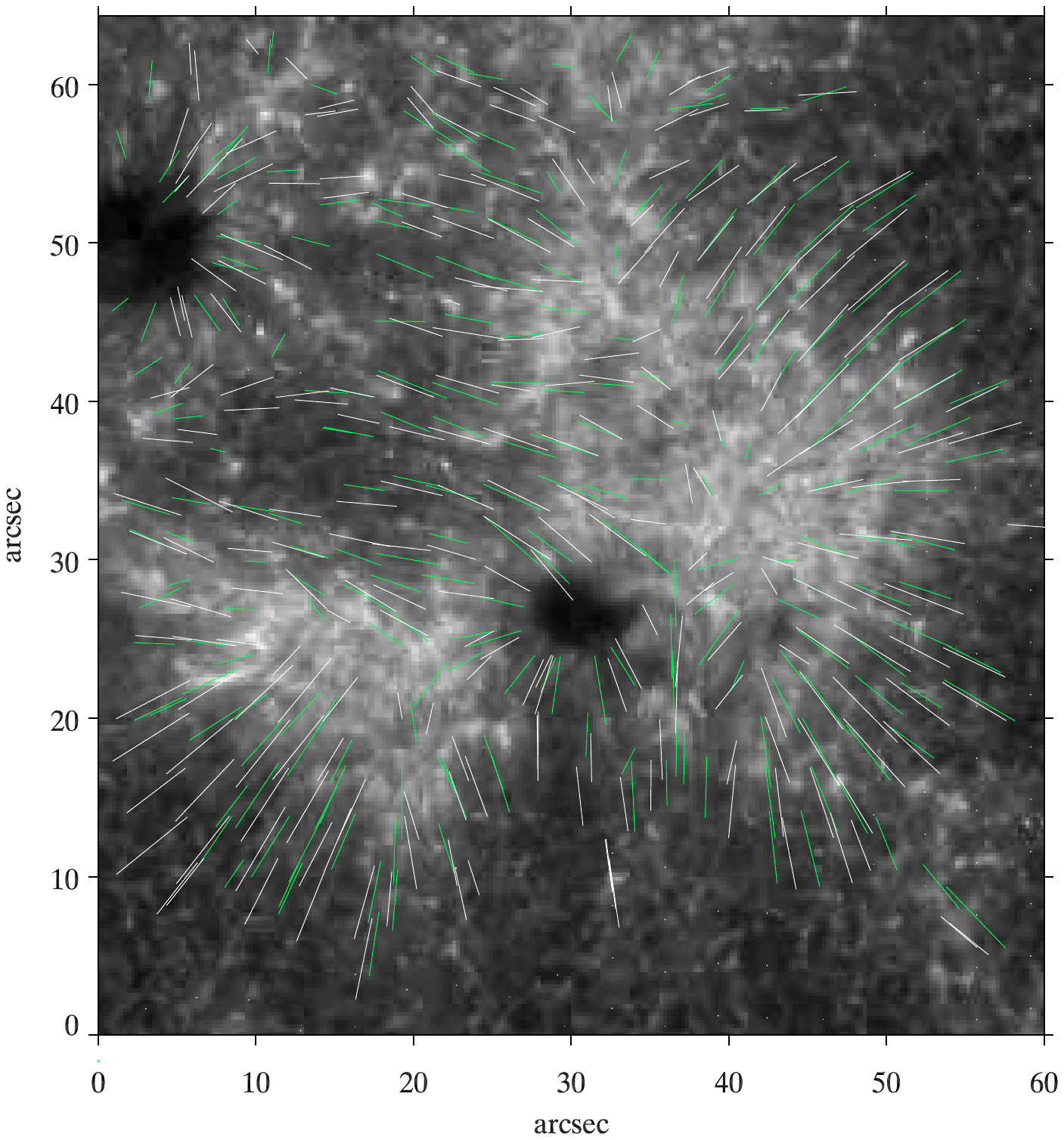}
\caption{Left: binary image of \ion{Ca}{II} K. Right: \ion{Ca}{II} K with over-plotted vectors
  showing prevalent direction of fibrils in the subregion. The green and
  white vectors are for images taken 36.5 minutes apart.}\label{fig:5}
\end{center}
\end{figure*}

The large-scale orientation pattern of the fibrils is shown in
Fig.~\ref{fig:5}. The top most image is a binary image made by unsharp
masking the \ion{Ca}{II} K image with a 12-pixel wide filter and then
thresholding it. The lower image shows the dominant fibril direction
in each subregion, as discussed in the previous section. The fibrils
extend (nearly radially) away from regions of strong magnetic field
towards the quiet Sun. The center of the largest plage (on upper right
of the central pore) shows rosette-like structures  with typical
  diameters of $\approx 10^{\prime \prime}$ (e.g., at $x=42
^{\prime \prime}$, $y=30 ^{\prime \prime}$). The plage is more
disorganized though even here most of the vectors are pointed towards
quiet Sun. The fibrils do not display apparent large loop-like structures in this
predominantly unipolar region observed at disk center. An exception is found in the upper left
corner between the two pores where faint, small curved-structures are
present. A granule surrounded by magnetic flux of the opposite
polarity than most of the FOV is located here ($x=13 ^{\prime
  \prime}$, $y=52 ^{\prime \prime}$). Since the region is near
  disk center, curvature in the fibrils is more likely due to the
  non-potentiality of the magnetic field configuration than to the
  loops being inclined with respect to the line of sight. Since the
central pore and the surrounding plage have the same polarity, no
loops connecting the two are present. The region in between and below
the plage and pore (located at $x=34 ^{\prime \prime}$, $y=24 ^{\prime
  \prime}$) has a very distinct cusp-like structure. There is a small
patch of opposite polarity to the lower right of the pore which some
of the cusp's fibrils connect to.

\subsection{Fibril types}

Four subregions (boxes in
Figs.~\ref{fig:1a},~\ref{fig:1c},~\ref{fig:1d} and \ref{fig:1b})
sampling different environments were chosen for a more detailed
study. Region A (plage edge and quiet Sun) has long fibrils extending
over several granules. Region B is located in the middle of the
plage. The cusp, i.e., fibrils originating from the pore and
surrounding plage, is shown in region C.  Small loop-structures
connecting opposite magnetic polarities are present in region D.

\subsubsection{Region A: Plage edge and quiet Sun}

\begin{figure*}[P]
\begin{center}
\includegraphics[width=12cm]{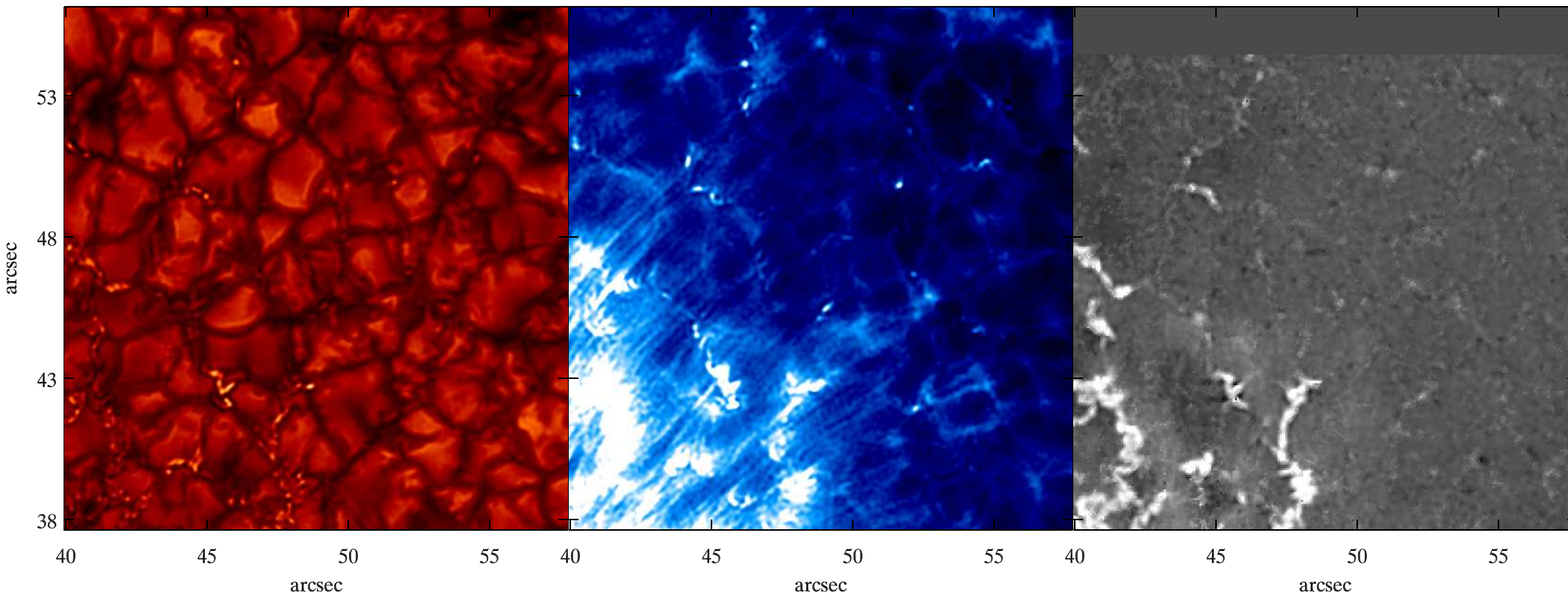}
\caption{Region A. From left to right: continuum intensity, \ion{Ca}{II} K intensity, longitudinal magnetic flux. The magnetic
  flux color scale goes from -800 to 1000 G.}\label{fig:A}
\end{center}
\end{figure*}

 Region A (Fig.~\ref{fig:A}) is located at the edge of the plage. The
 lower left corner in the continuum image is covered with filigree and
 many fibrils originate from there. The denser the
 filigree, the more jumbled and packed together the fibrils are. The
 long fibrils originating from close to the plage edge are oriented
 towards the quiet Sun where no large concentrations of magnetic flux
 are present and reversed granulation is the dominant intensity
 pattern. The long fibrils are nearly parallel to one another. They
 can extend over several granules and also over filigree that is the
 foot point of additional fibrils. Fibrils do not originate from all
 the bright points/filigree in the photosphere. In particular, the
 more isolated points in the quiet Sun are not associated with
 fibrils.

Closer inspection reveals that some of the apparent fibrils mirror the
shape and location of photospheric filigree. Given the large
contribution to the observed radiation from the photosphere (Sect. 2)
these may actually be photospheric filigree shining through.  Many, if
not all, of the fibrils with kinks are such filigree. The transition from
filigree to fibril is not an abrupt one: the photospheric filigree has
the same shape in the continuum and \ion{Ca}{II} K except that in
\ion{Ca}{II} K the filigree is often elongated/stretched towards the
quiet Sun as it transitions into fibrils.

\subsubsection{Region B: Strong plage}

\begin{figure*}[P]
\begin{center}
\includegraphics[width=12cm]{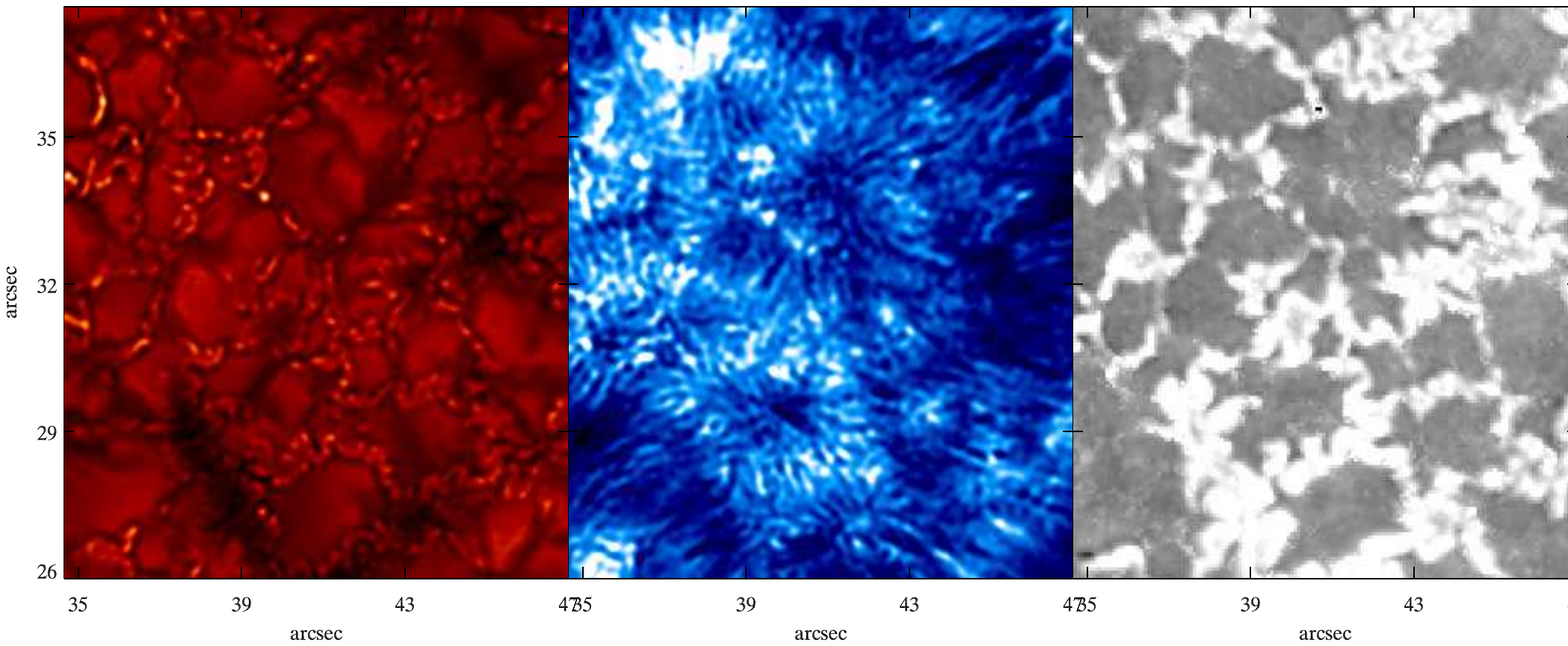}
\caption{Region B: As Fig.~\ref{fig:A}, but for region B.}\label{fig:B}
\end{center}
\end{figure*}

Region B (Fig.~\ref{fig:B}) is located in the strong plage. In the
photosphere the intergranular lanes are filled with filigree and
$\approx$kG-magnetic flux. The \ion{Ca}{II} K image is covered by a
carpet of short, stalky fibrils. The darker areas in the carpet, which
correspond to the centers of the rosettes, can be identified as
granule interiors. In the photospheric intensity and magnetic flux
rosette-like structures are formed at the intersection of three or
more granules whereas in the \ion{Ca}{II} K rosette-like structures
are formed in a granule that is surrounded at all sides by
intergranular lanes filled with filigree. Overlapping structures
complicate the identification of fibril endpoints, though the
fibrils clearly originate from the magnetic flux-filled intergranular
lanes. In places it appears as if the fibrils cross each other and are
in lying several layers. An example is located in the lower left corner of
Fig.~\ref{fig:B} (at $\approx$ $x=36 ^{\prime \prime}$, $y=28^{\prime
  \prime}$).

\subsubsection{Region C: Pore and plage}

\begin{figure*}[P]
\begin{center}
\includegraphics[width=12cm]{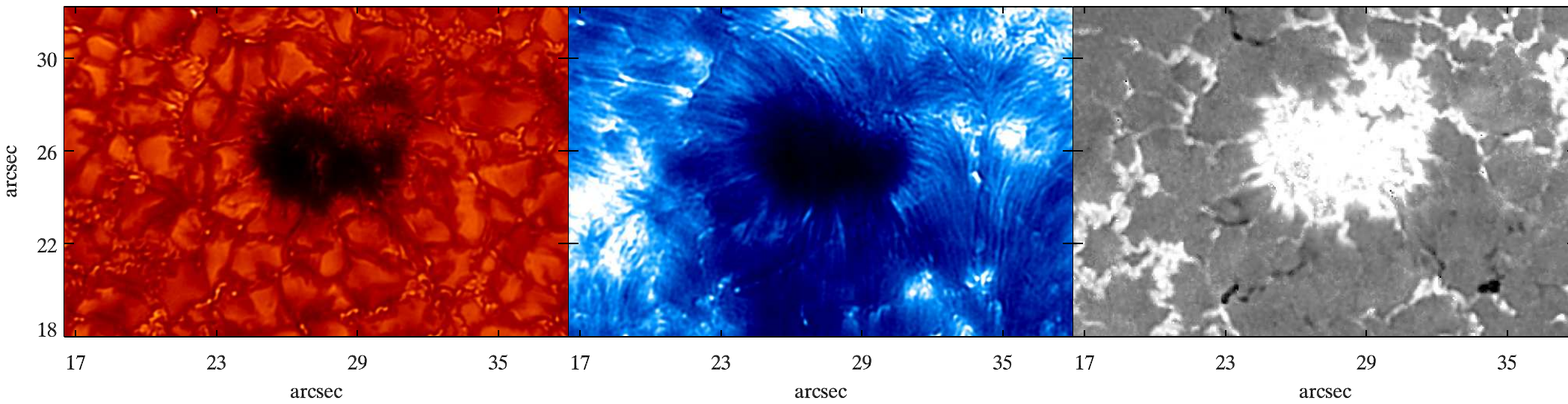}
\caption{Region C. As Fig.~\ref{fig:A}, but for region C.}\label{fig:C}
\end{center}
\end{figure*}

There is no large difference between the sizes of the pore (darkening)
in the continuum and \ion{Ca}{II} K (compare the corresponding frames
in Fig.~\ref{fig:C}). The small pore at the upper right of the large
pore is not well visible in \ion{Ca}{II} K, although it is associated
with a strong magnetic field (bottom panels in
Fig.~\ref{fig:C}). Instead, in \ion{Ca}{II} K the region is covered
with short fibrils. Fibrils originating from the pore form a structure
similar to a superpenumbra around the pore. The width of the
superpenumbra is comparable to the radius of the pore, $\approx$
$3^{\prime \prime}$.

Fibrils in Region C (Fig.~\ref{fig:C}) are oriented away from the pore
and plage. Since the two are of the same magnetic polarity, the
fibrils form a cusp-like structure (at [$x=34^{\prime \prime}$, $y=22-26^{\prime \prime}$]), i.e., they are deflected from the nearly
radial orientation to being mostly parallel to one another. A portion
of the fibrils involved in forming the cusp connect to the small patch of
opposite (negative) polarity below. The fibrils connecting to the opposite polarity have slightly different orientations and lengths. It appears as if the fibrils are not all in the same layer,
but instead the fibrils connecting to the opposite polarity are
lower in the atmosphere. There is another small negative polarity
patch at $\approx$ [$x=34^{\prime \prime}$, $y=30^{\prime \prime}$]
which appears to be an endpoint of some of the superpenumbral fibrils.

The large pore has a very clear superpenumbra in \ion{Ca}{II} K
emission except in the upper right corner where the pore is in
direct contact with the plage. In the photosphere the granulation
pattern in the corner is disrupted. The corner does not have a clear
superpenumbra but instead it is more similar to the dense plage area
than the rest of the pore, i.e., the \ion{Ca}{II} K fibrils are more
carpet-like. The magnetic flux in the plage is not much lower than in
the pore. This may, however, be due to the high level of straylight which
dominates in the umbra. The rest of the pore-superpenumbra boundary serves as the foot point to fibrils that are oriented nearly radially away from the
pore. They can extend outside the superpenumbra indicating
that some of the fibrils are located at a different height in the atmosphere than
the bulk of the superpenumbra. Since the region is unipolar, the superpenumbra  cannot connect back to the photosphere but instead must expand higher up and disappear from view. In contrast, some of the fibrils remain visible beyond the superpenumbra.

\subsubsection{Region D: Bipolar weak plage}

\begin{figure*}[P]
\begin{center}
\includegraphics[width=12cm]{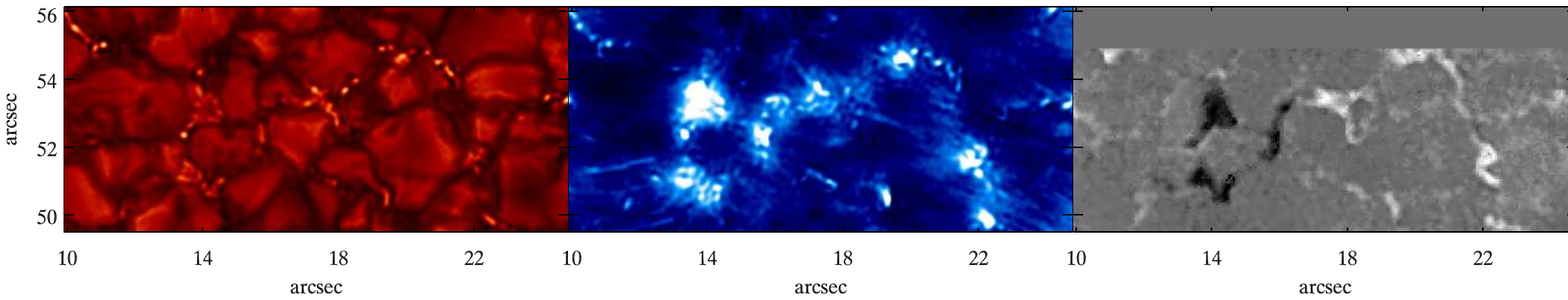}
\caption{Region D. As Fig.~\ref{fig:A}, but for region D.}\label{fig:D}
\end{center}
\end{figure*}

Region D (Fig.~\ref{fig:D}) is located between the two pores. Some of the intergranular lanes are filled with filigree that have the same
shape in the continuum and \ion{Ca}{II} K images. The intergranular
lanes surrounding one of the central granules are filled with filigree
which coincides with magnetic flux of the opposite polarity than most of the FOV. The opposite polarity field is the foot point of
fibrils that extend toward the directions of both pores and form small loop-structures
which are best seen in unsharp masked images. The large scale
structure of the fibrils is more disorganized than in other regions
with a moderate (though still less than in the plage) amount of
magnetic flux. Most of the surface in region D is covered by
fibrils. This is best seen in Fig.~\ref{fig:1b}. In regions with no filigree the fibrils are parallel to one
another whereas in the filigree-filled regions they form rosettes
similar (though not as dense) to the plage (region B).

It should be noted that in the current dataset fibrils cannot be used as a reliable means of determining the magnetic connectivity of the photosphere. For example, in the right side of regions D some of the long, clear fibrils appear as if they are connecting two features. However, inspection of the magnetic field reveals that these two features are of the same polarity and thus cannot be connected magnetically.
 
\section{Fibril dimensions}

\begin{figure*}[P]
\begin{center}
\includegraphics[width=6cm]{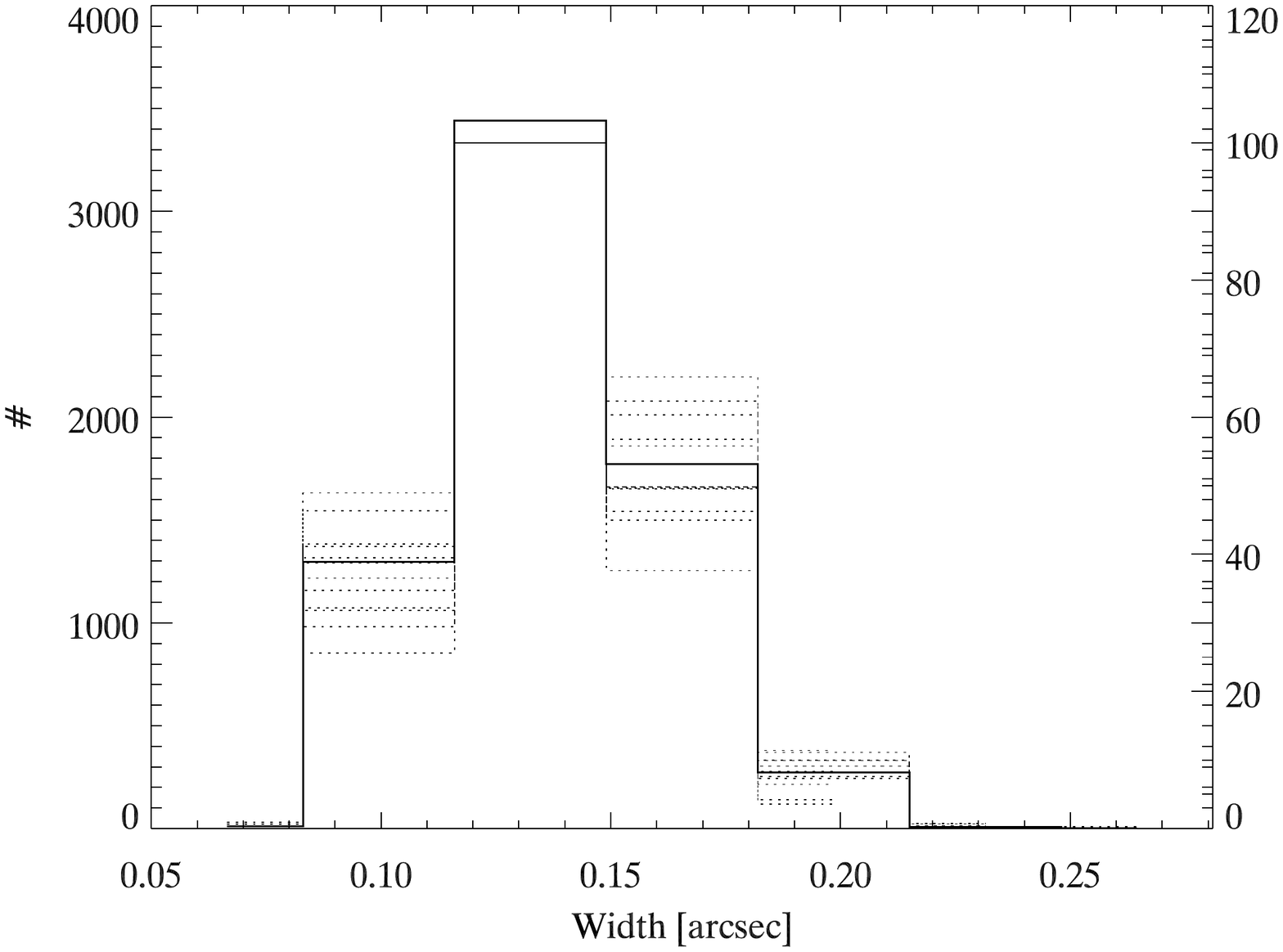}
\caption{Histogram of automatically detected fibril widths. The solid
  line is the average histogram for all 12 images, dotted lines for the individual images.
}\label{fig:4a}
\end{center}
\end{figure*}

The histogram of automatically detected bright fibril segment widths
(Fig.~\ref{fig:4a}) peaks at 0.11-0.14 arcseconds (i.e., only slightly
above the diffraction limit, indicating that the actual widths may
be below 0.1 arcseconds). The distribution goes down to values below
the diffraction limit and up to 0.2 arcseconds. No correlation is found between the fibril
intensity and width. The lengths of the segments are comparable to the
length scale of the reversed granulation (which is the underlying
intensity pattern): most of the segments are less than
1.5 arcseconds long. The lower limit of the
lengths of the fibril segments is set by the criterion used in
identification (0.825 arcseconds). 

\begin{figure*}[P]
\begin{center}
\includegraphics[width=14cm]{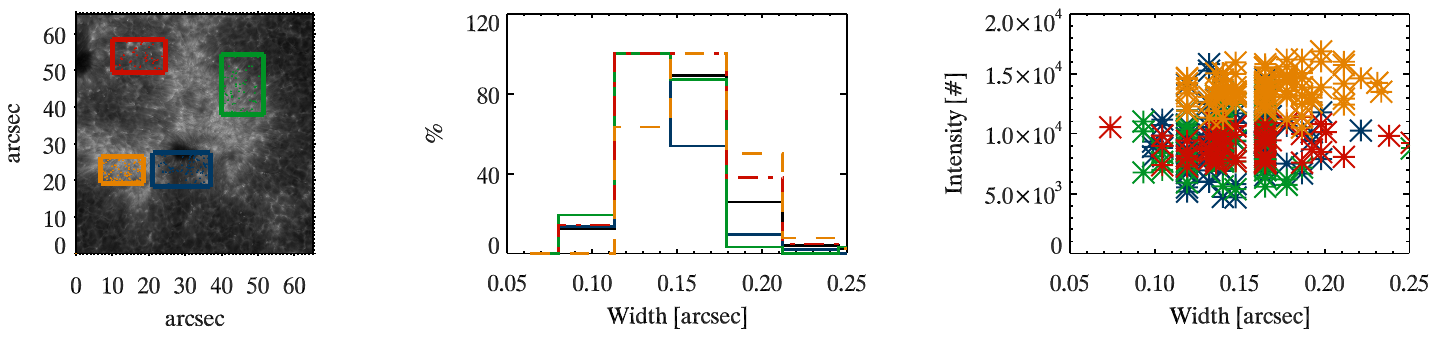}
\caption{Manually measured fibril widths. First image: colored boxes
  in the Ca K intensity image show the locations of the four regions
  where the fibrils with manually measured widths are located. Second image: histograms of fibril
  widths. Black line is for all four regions, colored lines correspond
  to the four regions marked in the first image. Third image: scatter
  plots of fibril widths and intensity. Color coding same as in
  histograms.  }\label{fig:4b}
\end{center}
\end{figure*}

The manually determined widths (Fig.~\ref{fig:4b}) are very similar to
the automatically detected widths: most fibrils are between 0.11 and
0.14 arcseconds wide. There is a pattern in the spatial distribution
of the widths: a larger portion of fibrils with widths below 0.14 and
fewer wide fibrils are found in region C (pore). Most of the fibrils
in region A are between 0.11 and 0.18 arcseconds wide. Region D (weak
bipolar plage) resembles region A except for the larger portion of widths
larger than 0.18 arcseconds. Region B (strong plage) has a wide
distribution and fibrils with widths up to and over 0.2 arcseconds are
more common than in the other regions. Again, no correlation is found
between the widths and the fibril intensities. However, the mean value
of the widths in a region increases with increasing mean intensity
(which in turn is correlated with the magnetic flux) of the whole
region (not mean intensity of the fibrils), i.e., the fibrils are on average thicker in the plage than in the quiet Sun. If this effect is related to the width of the spectral filter employed for the observations is unclear: in the quiet Sun chromospheric layers contribute little to the observed intensity, so that fewer fibrils are visible and only the brighter central spines of the fibrils are seen. In regions with more magnetic flux more chromospheric contribution is obtained and also fainter parts of the fibrils are seen making them appear broader.

Interference from the underlying reversed granulation and lack of
well defined fibril ends make measuring lengths difficult, and only a
rough measurement was done by hand. Fibrils extending over quiet Sun
are longest (from 6 up-to 10 arcseconds) and fibrils in regions with more magnetic flux and stronger unipolar
crowding are significantly shorter (1-1.5 arcseconds).
 
\section{Temporal evolution}

The main intensity variation in the \ion{Ca}{II} K images is in the
middle photosphere and is due to reversed granulation. Because of this
it is difficult to distinguish changes in fibril intensity from
changes in the background intensity. This in turn makes it difficult
to follow individual fibrils from one frame to the next.

A variety of dynamic phenomena such as splitting, merging, fading, brightening
and elongation are seen in the fibrils. The large scale structure is very stable throughout the observing
sequence and does not change significantly over 36.5 minutes (compare the white and green lines in
Fig.~\ref{fig:5}). The mean change in orientation (measured as the
difference in angle of the white and green lines in each subregion) is
only 1.6 $\deg$ and the standard deviation is 13.9 $\deg$, i.e.,
there is no large scale change in the fibril orientation. This means
that the individual bright \ion{Ca}{II} K fibrils are replaced by new ones
with very similar orientation and general appearance or,
alternatively, the fibrils live longer than the duration of the timeseries while the \ion{Ca}{II} K emission disappears and reappears in the same
location. 

 Small-scale evolution is
fastest in the strong plage regions while the bipolar region (region
D) undergoes the fastest larger scale changes. This is because the
small patch of opposite polarity magnetic field changes significantly during the observing
sequence. The unipolar regions (e.g., region A) with fibrils extending
over several granules and the pore-plage interface (region C) are very
stable. The fibril dynamics may be driven by motions in the
photosphere. If the granulation and its associated filigree are evolving fast,
then the fibrils are more dynamic. This is clearly seen in movies made
of the photospheric and chromospheric observations, when watched in
parallel.

\begin{figure*}[P]
\begin{center}
\includegraphics[width=14cm]{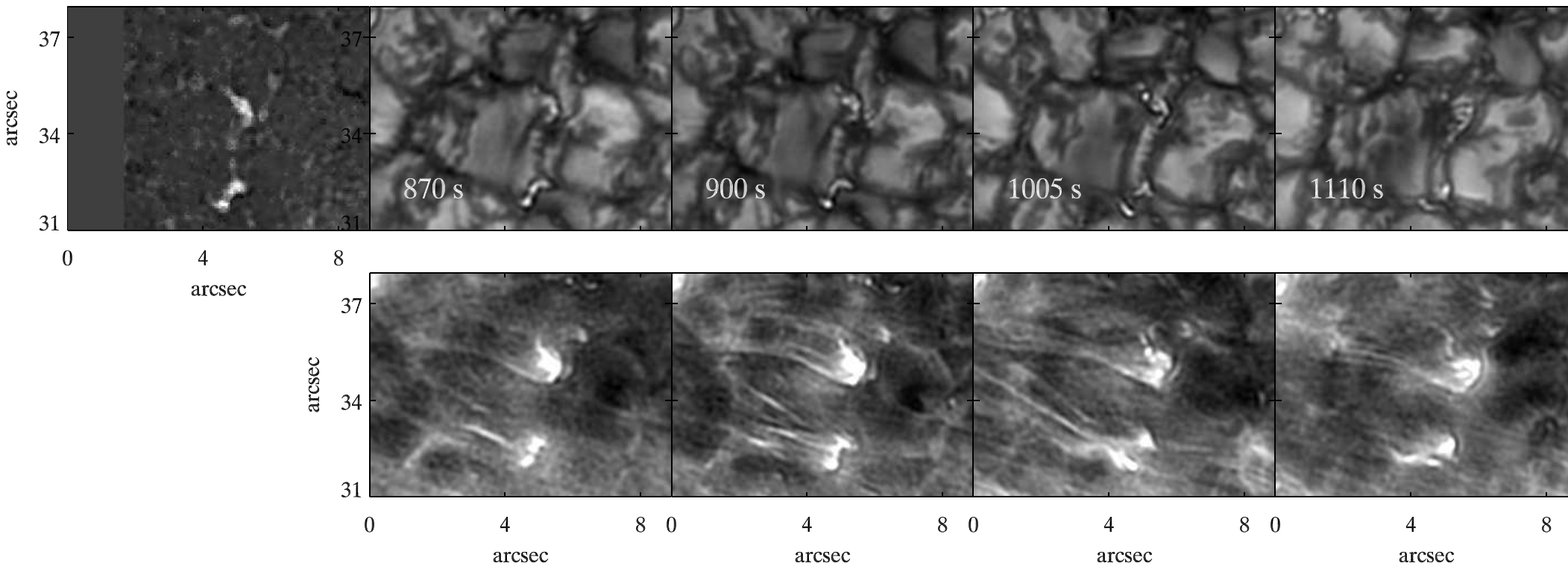}
\caption{$1^{st}$ row: longitudinal magnetic flux and
  continuum snapshots of the same region. Time relative to the beginning of the observing sequence of each snapshot is shown
  in the corner of each image. $2^{nd}$ row: \ion{Ca}{II} K snapshots
  corresponding to the continuum snapshots. Note that the images are irregularly spaced in time. }\label{fig:6a}
\end{center}
\end{figure*}

An example of temporal evolution of photospheric filigree and individual fibrils associated with the filigree is shown in Fig.~\ref{fig:6a}. The region is located close to the edge of the plage. The bright points and filigree are most likely foot points of the
\ion{Ca}{II} K fibrils seen in the snapshots. The appearance of the photospheric filigree changes during the sequence and at the same time the \ion{Ca}{II} K fibrils become
gradually brighter and more elongated. A rough estimate of the
elongation speed is $\approx$ 20 km $\mathrm s^{-1}$.  After a couple of minutes the elongated fibrils fade away. Other examples
of elongating fibrils with similar speeds are found at the edges of
the plage.

\begin{figure*}[P]
\begin{center}
\includegraphics[width=14cm]{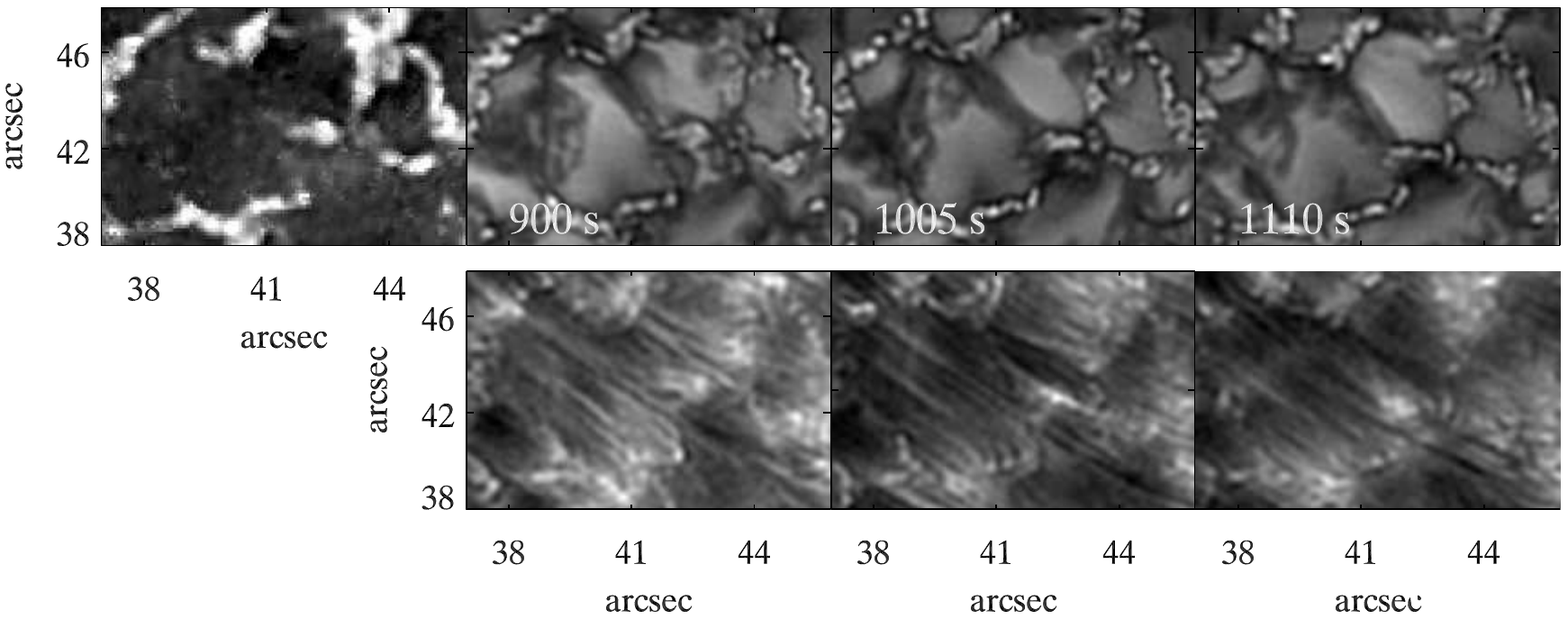}
\caption{$1^{st}$ row: longitudinal magnetic flux and continuum snap shots of the same region. Time of each image is shown above each image. . $2^{nd}$ row:: \ion{Ca}{II} K snapshots taken at the same time as the continuum images.}\label{fig:6b}
\end{center}
\end{figure*}

Individual fibrils in the plage undergo changes in visibility, but overall the structure changes only little during 3.5 minutes
(Fig.~\ref{fig:6b}). The largest changes are in the underlying granulation
pattern which alters the visibility of the fibrils.  Some of the fibrils
merge and split intermittently. Their orientation can also change
slightly. In contrast, the long fibrils extending over the quiet Sun
tend to fade away (instead of merging and splitting) and do not show
changes in orientation.  A movie of the \ion{Ca}{II} K images around
the plage region (on the left in the FOV) shows an outward moving pattern along the fibrils from the plage towards the quiet
Sun (no identification of this pattern with photospheric structures can be made). Another type of lateral movement is occasional bright points that
appear to be moving along the fibrils. However, all of these points
can be identified as photospheric bright points shining through. It is
not possible to say if the fibrils sway in the lateral direction
because of the jitter caused by the varying seeing conditions and
the slow cadence (15 s) of the data. In general, the average lifetimes of the
fibrils are at least 3-4 minutes which is also the time that the
individual fibrils can be tracked reliably. It should be noted that
this timescale is similar to the timescale of the underlying
granulation.

The general structure of the cusp (region C) remains the same
throughout the time series.  The patch of opposite magnetic polarity
into which the fibrils connect does not disappear during the observing
sequence and the individual fibrils, especially the ones connecting to
the opposite polarity patch, are long-lived. The time evolution is
smooth (they merge, split or fade away) and it is not straightforward
to say exactly when a feature is born or dies away. Brightening and
darkening of individual points or ``crinkles'' in the filigree are not
well correlated in the photosphere and chromosphere.

\subsection{Power spectra}

Power maps of the \ion{Ca}{II} K and continuum intensity are plotted in
Fig.~\ref{fig:7a}. Shown is a subregion located on the right side of
the FOV. The three different frequency ranges (evolutionary,
evanescent and propagating) are discussed individually in the
following subsections. The evolutionary frequency range encompasses
frequencies from 0.4 to 1.6 mHz (periods 10.4 to 41.7 min), evanescent
frequency range 2.4-4 mHz (4.2 to 6.9 min) and propagating frequency
range 5.2-8 mHz (2.1 to 3.2 min). 

\begin{table}
\caption{Power in regions I, II and III}             
\label{table:power}      
\centering                          
\begin{tabular}{c c c c}        
\hline\hline                 
Frequency range & region I: & region II: & region III: \\
mHz & Bright plage & Intermediate plage &  Quiet Sun \\    
\hline                        
Continuum & & & \\
0.4-1.6  & 0.69 & 0.85 & 1.17 \\
2.4-4    & 0.82 & 0.90 & 1.11 \\
5.2-8    & 1.02 & 0.97 & 1.01 \\
Ca II K     & & & \\
0.4-1.6  & 0.73 & 1.02 & 1.06 \\
2.4-4    & 1.12 & 0.85 & 1.06 \\
5.2-8    & 0.71 & 0.69 & 1.27 \\

\hline                                   
\end{tabular}
\end{table}

\subsubsection{Evolutionary range $<$ 1.6 mHz}

In order to compare the power in regions of different chromospheric brightness the FOV of Fig.~\ref{fig:7a} is divided into three subregions based on the
average \ion{Ca}{II} K brightness (bright plage: region I in Fig.~\ref{fig:7a}, intermediate plage: region II
and quiet Sun: region III). Table \ref{table:power} shows the amount of power in regions
 I, II and III relative to the mean
power of the entire FOV in each frequency range. In both, the \ion{Ca}{II} K and continuum, the
brightest plage, which also corresponds to the strongest magnetic
flux, has clearly less power in the evolutionary frequency range than the surroundings. There is no
difference in \ion{Ca}{II} K between regions II and III, but in the
continuum there is more power in the quiet Sun (region III) than the
weak plage (region II).

The patches of enhanced photospheric power in the left side of the
plage (Fig.~\ref{fig:7a}) correspond to the micropores which start to slowly fade away
towards the end of the time series. The power enhancement due to the
micropores is not visible in the \ion{Ca}{II} K image because the
plage is dominated by chromospheric emission from the fibrils.

The fibril structure is clearly visible in the low frequencies of the
\ion{Ca}{II} K intensity power. The individual fibrils are seen as
faint bright features which extend beyond the bright plage. This is
worth noting, since they are not apparent in the averaged brightness
image (top right panel of Fig.~\ref{fig:7a}). The enhanced power may correspond to the gradual brightening and
fading away of fibrils. Other regions with increased power are located
in the plage at the foot points of the fibrils and in the quiet Sun,
where they often correspond to brighter than average intensity and
more isolated filigree. The isolated filigree does not appear as
bright in the continuum intensity and, consequently, the regions do
not exhibit enhanced photospheric power. A local power increase is
visible in the superpenumbra of the pore. This corresponds to a sudden
brightening seen in Ca II K towards the end of the time series. During
the brightening the granulation pattern changes abruptly and a small enhancement is also seen in the continuum power.

\subsubsection{Evanescent frequencies 2.4-4 mHz}

\ion{Ca}{II} K power in the evanescent frequencies is suppressed at
the edges of the plage (region II) whereas the brightest plage (region
I) shows no suppression. In fact, slightly enhanced power in \ion{Ca}{II} K is
seen by the foot points of the power fibrils. In contrast, region
I does not have enhanced power in the photosphere, instead the brightest
plage has the least power in this frequency range. This leads to the
conclusion that the enhanced power in the bright plage is
chromospheric, not photospheric. In the bright plage individual power
fibrils are clearly visible in the \ion{Ca}{II} K power images
(Fig.~\ref{fig:7b}). In fact, they are far better visible in power than in
average brightness. Regions of enhanced \ion{Ca}{II} K power in the
quiet Sun are likely due to the photospheric contribution.

\subsubsection{Propagating frequencies 5.2-8 mHz}

\begin{figure*}[P]
\begin{center}
\includegraphics[width=14cm]{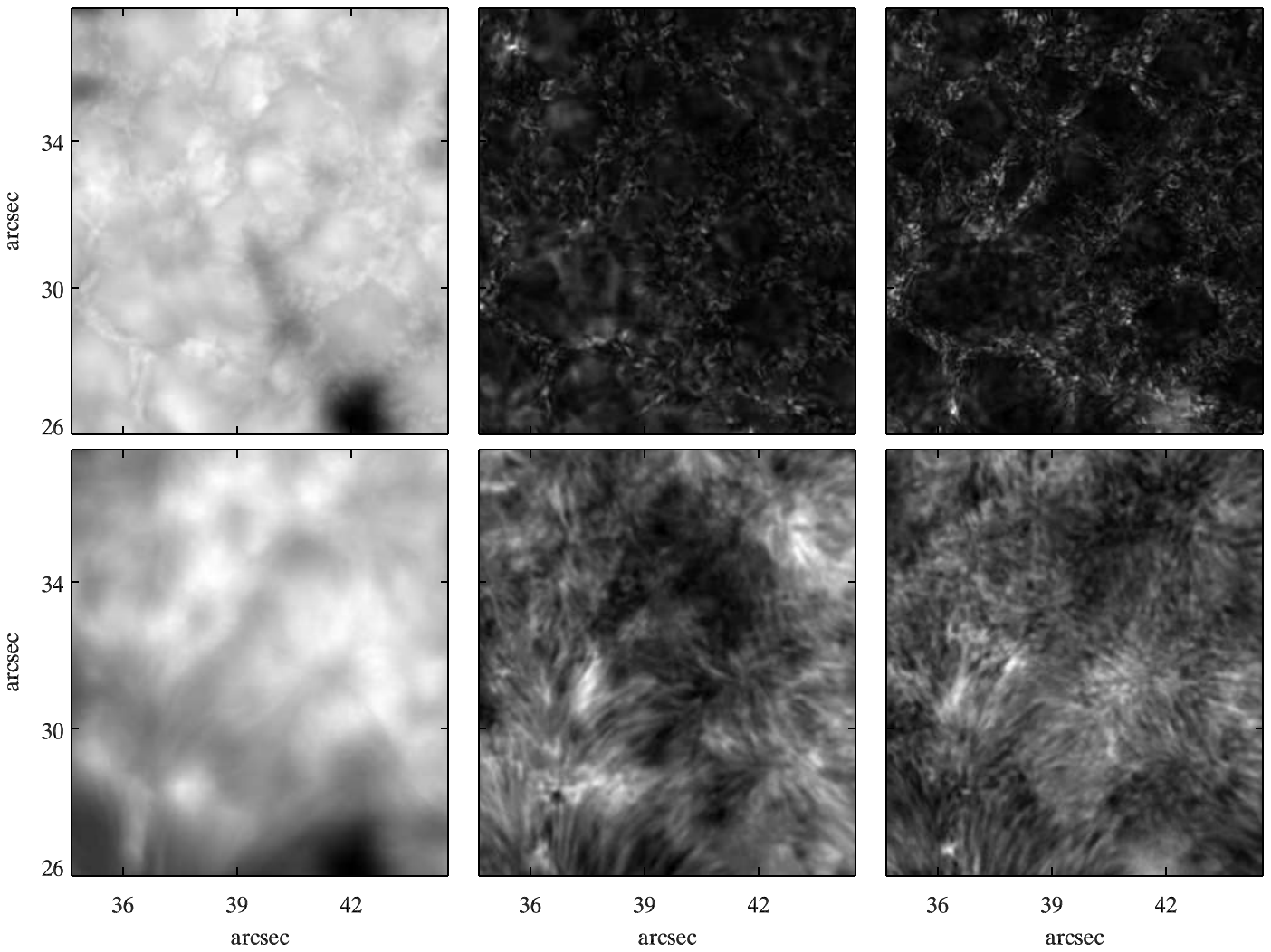}
\caption{Power map of strong plage. $1^{st}$ row, $1^{st}$ image: average
  continuum intensity. $2^{nd}$ image: continuum power 2.6-4
  mHz. $3^{rd}$ image: continuum power 5.5-8 mHz. $2^{nd}$ row as
  $1^{st}$ but for \ion{Ca}{II} K.} \label{fig:7b}
\end{center}
\end{figure*}

The third frequency range, above the acoustic cutoff
frequency, is where the internetwork chromospheric power typically peaks. In the photosphere, the power in this range shows a rapid drop relative to the power of the dominant five-minute periods. Consequently, the photosphere has overall little power
in this range and the spatial pattern in the power map is related to
intergranular lanes. In fact, no significant difference is seen
between the three regions in the photosphere. Chromospheric power is suppressed both in the
bright plage (region I) and the weaker plage (region II). The region of suppressed power extends beyond region II well into region III. Furthermore,
individual power fibrils at the plage edge appear as darker structures,
i.e., the intensity of the fibrils varies less than the
surrounding intensity. Also in strong plage, fibril-like structures are well visible in this frequency range (Fig.~\ref{fig:7b}). Unlike in the evanescent regime, the fibril
foot points do not show up as regions of greatly enhanced power. The
power enhancement in the propagating range is only modest at the
fibril foot points. In contrast, significantly more power is seen in
the quiet Sun (region III) in \ion{Ca}~{II}~K.

\begin{figure*}[P]
\begin{center}
\includegraphics[width=14cm]{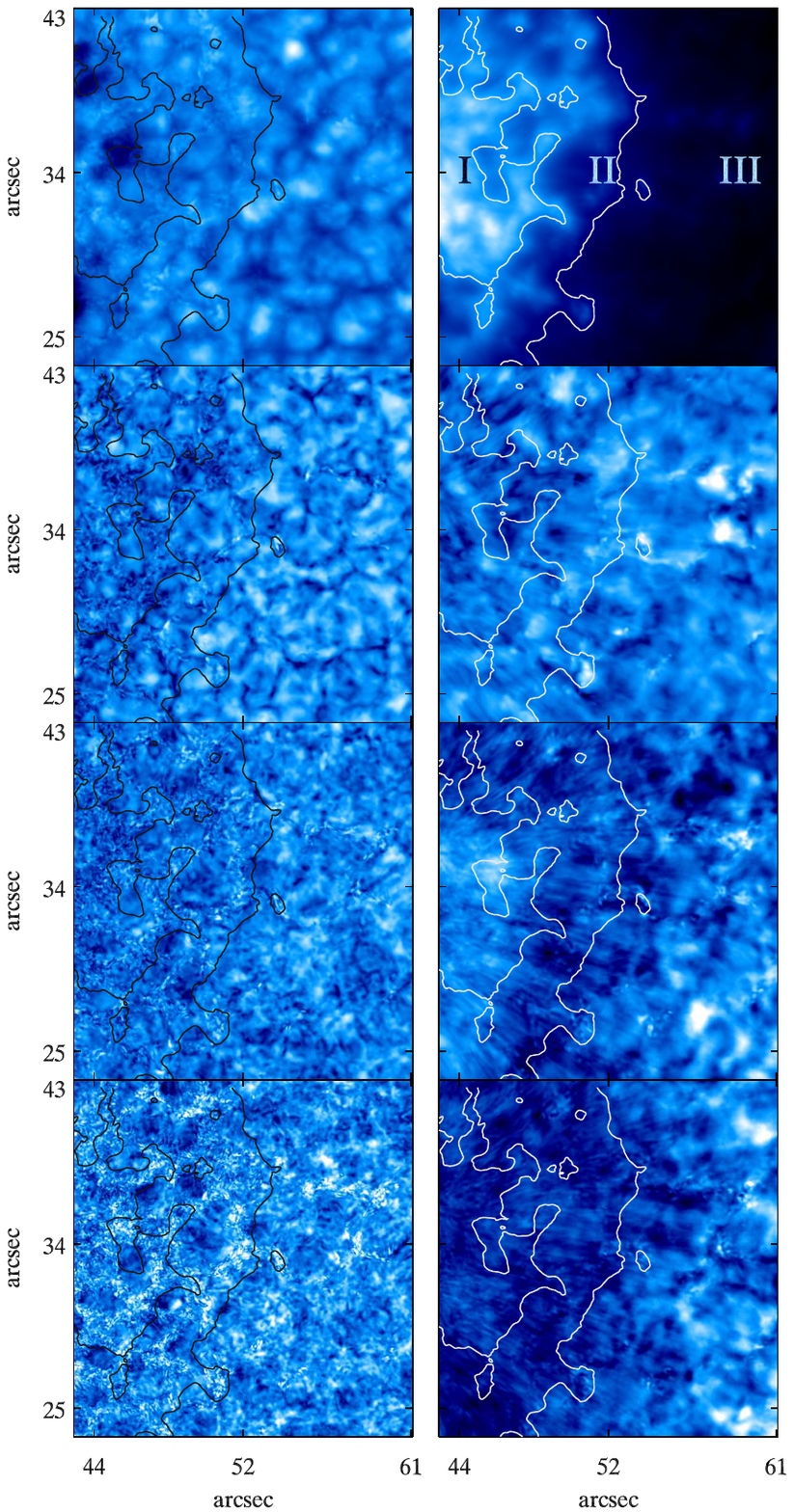}
\caption{Power maps of a subregion in the FOV.  $1^{st}$
  column, $1^{st}$ row: average of all continuum intensity images with
  good seeing.  The contours outline strong plage (left contour region I) and weak plage (right contour, region II) relative to quiet Sun (region III) in the \ion{Ca}{II} K
  emission. $2^{nd}$ row: continuum power at 0.4-1.6 mHz. $3^{rd}$
  row: continuum power 2.4-4 mHz. $4^{th}$ row: continuum power 5.2-8
  mHz. Second column: as first row except for \ion{Ca}{II} K. The
  color scales are chosen to highlight differences in the images.}\label{fig:7a}
\end{center}
\end{figure*}

\section{Discussion}

The current data show that fibrils are a prominent feature in
\ion{Ca}{II} K. The presence of fibrils indicates that a significant fraction of the signal through this filter is coming from chromospheric structures in these regions. The single most important result of this investigation is that the fibrils in \ion{Ca}{II} K are even more abundant and finely structured than ever observed. Prior observations of \ion{Ca}{II} K \citep{Beckers1968, Zirin1974, Marsh1976, Rutten2007} have shown fibrils but not as clearly and in as much detail as in the the current data. The fibrils have not been observed as conspicuously
before because of a variety of reasons. Only the \ion{Ca}{II} K line
core is purely chromospheric, the wings are formed in the
photosphere. Since the line core is very dark, relatively few photons
originate from the chromosphere.  This means that a 1.5 \AA\ filter is probably close to the maximum width at which fibrils outside the strongest plage can be seen. Another factor is the very small apparent fibril width through this broad filter, which leads to very
high spatial resolution (and short exposure time) requirements. For ground-based observations these are combined with the need of image reconstruction methods.

Similar structures as the \ion{Ca}{II} K fibrils have been observed
with the Dutch Open Telescope using a 1 \AA\ wide \ion{Ca}{II} H
filter \citep{DOT2004, Rutten2007}. These bright structures called straws (\citealt{Rutten2007})
originate from magnetic regions such as active network and plage. They
usually coincide with the lower ends of bright H$\alpha$\ fibrils.  The
straws are qualitatively very similar to the bright \ion{Ca}{II} K fibrils,
especially in the small patch of network present in the lower right
corner of the FOV. 

In the observations the bright \ion{Ca}{II} K fibrils are seen everywhere in
the data where an increased fraction of the signal is coming from the
chromosphere. We cannot rule out that if the signal was not dominated by the photospheric
contribution in the quiet Sun,  the
fibrils would be as ubiquitous there as in the vicinity of the
plage. In fact, this is rather likely. The fibrils are structured by the underlying magnetic field and
the bright endpoints are clearly cospatial with the magnetic
concentrations in the photosphere. This is not as obvious in the strong plage where the density of the fibrils is high enough to make identifying the endpoints difficult. In the plage where there is more
magnetic flux and due to unipolar crowding the field is more vertical,
the fibrils are more vertical and form a carpet covering the surface. Towards the
edge of the plage the fibrils become longer and more organized. At the
edge they extend radially away from the magnetic
concentrations and are nearly parallel to one another. The long
fibrils extend over several granules and form a multi-layer (in the
sense that not all the fibrils are at the same height) canopy over the outer part of the plage and the quiet Sun. 

Since the fibrils appear bright in the \ion{Ca}{II} K emission, this indicates a
temperature enhancement in the chromospheric layers where the
\ion{Ca}{II} K line is formed in (e.g., ~Fig.\ref{fig:0}). Why the
general structure in the \ion{Ca}{II} K emission consists of fibrils instead of a more uniform, homogeneous structure
is an interesting question. The fibrils are clearly closely related to
the magnetic field and it seems plausible that they outline the
general magnetic field structure as proposed by \cite{Veeder+Zirin1970}. Perhaps the lower boundary of the field is similar to a corrugated sheet
where the \ion{Ca}{II} K fibrils are material with sufficient opacity in the dips of the sheet
 coinciding with the formation range of the line. Another possible explanation is that the magnetic canopy itself is rather homogeneous and some of the magnetic
field lines have an enhanced optical
depth because they are either more heated or are filled with more
material than other field lines. This would result in a visible
canopy-structure formed of individual fibrils. Non-uniform heating of
the canopy may be due to reconnection or waves. If the differential
heating mechanism is waves, the question arises why are the waves
excited so locally or why are they channeled only by a portion of the
magnetic field lines.

Another question is why do the fibrils remain narrow, coherent
structures when the magnetic field, and hence also the fibrils, are
expected to expand with height? The combination of noise and threshold
may partially explain this: at the foot points the full width of the
fibrils is seen, but with increasing distance the fibrils become
darker and only the bright core is visible while the fainter edges are
drowned in the background noise. Since the contribution function of
the filter used in the observations is wide, the background noise
(i.e., photospheric contribution) is strong and only the brightest
central part of the fibrils is above the noise level. This would also explain why the quiet Sun fibrils are narrower than the fibrils in the plage. 

The fibrils extending over the quiet Sun fade gently out of view. This
could be because as the magnetic field expands it moves to higher
atmospheric layers and, consequently, outside the range of formation
of the \ion{Ca}{II} K line. Alternatively, the heating mechanism may
be more concentrated at the foot points of the fibrils.

Regardless of the speculations above, the \ion{Ca}{II} K fibrils
support the presence of a hot magnetic canopy (e.g,
\citealt{Solanki+others1991}, \citealt{Solanki+others1994},
\citealt{Ayres2002}). Furthermore, the coherence of the fibrils lends
support to the classical picture of the magnetic canopy (e.g,
\citealt{Jones+Giovanelli1982}, \citealt{Solanki+Steiner1990},
\citealt{Bianda+others1998}) in contrast to the picture where magnetic
concentrations in the internetwork disrupt the formation of a coherent
large-scale canopy \citep{Schrijver+Title2003}. These two scenarios need not necessarily exclude each other entirely. It may be that the small scale magnetic loops and structures are mostly restricted to the photosphere and/or low chromosphere whereas the canopy is located higher up in the chromosphere. It may also be that the truly quiet Sun lacks a canopy while in other more active regions, such as in the vicinity of chromospheric network, a canopy is present.

The power maps of the \ion{Ca}{II}~K intensity are very similar to the
power maps of \ion{Ca}{II} 8542 \AA\ velocity studied by
\cite{Vecchio+others2007}. The role of the fibrils in the dynamics is
two-fold: firstly, they are associated with the channeling of low
(below acoustic cutoff) frequency waves propagating into the
chromosphere (e.g., \citealt{Jefferies+others2006}). An inclined magnetic field
lowers the effective gravity and also the acoustic cutoff frequency of
a wave propagating along the field line \citep{Bel+Leroy1977}. Indications of this effect are seen in
the power maps (Fig.~\ref{fig:7a}) where the foot points of the
fibrils appear as regions of enhanced power in the frequency regime
normally evanescent in the chromosphere. The
more inclined the field is, the more the acoustic cutoff frequency is
lowered. This may explain the long periods in the nearly horizontal
fibrils originating from the plage edge. Similar oscillation
properties have been observed in e.~g.~ H$\alpha$\ fibrils: long
periods (up to ten minutes) are found in the nearly horizontal mottles
while the more vertical dynamic fibrils are dominated by periods of
four to six minutes \citep{DePontieu+others2007b}.

In the present observations the largest enhancement of power at
frequencies below the acoustic cutoff is not in the region with
strongly inclined field at the edge of the plage, but rather in the
strong plage, where the fields are expected to be more vertical. This
is not in agreement with the classical picture of better propagation
in more inclined fields. Since the photospheric power is not enhanced in the region, the chromospheric enhancement cannot be due to a more efficient
excitement of these waves. Consequently, either such oscillations are
better visible in more vertical magnetic structures, or there are as
yet unknown reasons for their more efficient propagation along nearly
vertical fields. The former may be partly explained if at the edge of
the plage region the lower parts of the atmosphere contribute
significantly to the observed signal, i.e., there is less opacity in
the fibrils than the underlying atmosphere and, consequently, we
mostly observe the suppression of oscillations from below instead of
wave propagation along the fibrils. In addition, if the wavelength of
the oscillations is significantly longer than the fibril width, then
there is only little change in optical depth due to the wave in nearly
horizontal fibrils. In contrast, in the more vertical fibrils the wave
changes the properties of the fibril over the full optical depth range
of the line formation. If the oscillations are guided by magnetic fields, then the amount of apparent power may also depend on the number of bright points. The greater number of bright points in region I with respect to region II would then imply a larger leakage, and hence also power, in region I.  

Furthermore, because of the wide contribution function of the filter,
the signal is entirely dominated by the photosphere unless there is additional
heating present in the chromosphere. If such a heating mechanism
operates mostly in the vicinity of strong flux concentrations, we
would not expect to see the more spatially sparse horizontal fibrils
as clearly as the fibrils in the strong plage. One can also speculate
whether the waves loose some of their energy through mode conversion when
they encounter a large (compared to the wavelength) magnetic field
(i.e. the wave-guide) curvature as the field becomes more horizontal
at the edge of the plage. This would be similar to what is seen in
simulations of coronal loops, where curved structures lead to increased leakage
of both sausage and kink waves \citep{Smith+others1997,
  Selwa+others2007}. Yet another alternative explanation for the lowered  acoustic cutoff frequency involves radiative cooling. A loose reasoning can be made where the
more inhomogeneous plage region has a shorter radiative cooling time
which in turn leads to a decreased acoustic cutoff frequency \citep{Centeno+others2006}.

The second role the fibrils have in the dynamics is that they act as a
canopy over the quiet Sun. The canopy
suppresses oscillations from below as is seen in the power map of the
propagating frequency regime (last panel in
Fig.~\ref{fig:7a}). Recently observational evidence supporting the
canopy scenario has increased significantly, e.g.,
\cite{Vecchio+others2007} find evidence for it in
\ion{Ca}{II} 8542 \AA\ data and \cite{Holzreuter+Stenflo2007} in
scattering polarization data. Suppression of oscillations in the
vicinity of magnetic network was reported by \cite{McIntosh+Judge2001}
in SOHO/SUMER data. They referred to the phenomenon as ``magnetic
shadows''. Magnetic shadows in \ion{Ca}{II}~K have been reported before by \cite{Tritschler+others2007, Woeger+others2006, Reardon+others2007}. Both the canopy and magnetic shadows are most likely
manifestations of the same phenomenon: inclined magnetic fields above
quiet Sun regions altering the propagation of waves \citep{McIntosh+Judge2001, Vecchio+others2007}. The current high
spatial resolution data demonstrate that the canopy is very
inhomogeneous and composed
of structures with sizes near the currently accessible spatial
resolution. This is not obvious in observations with lower spatial
resolution.

The effect of the canopy in the very quiet Sun cannot be determined since
the signal in these regions is probably coming from much lower (i.e., the chromospheric
contribution is very small) and also the canopy (in the current
context defined as strongly inclined fibrils that can alter the
dynamics of upward propagating waves) is probably located higher
up. The \ion{Ca}{II}~K fibrils are not visible above
the quiet Sun: the magnetic field that outlines
the fibrils may be above the formation height of the line or the
fibrils may not have sufficient density and/or temperature above the
quiet Sun. Simultaneous observations with H$\alpha$\ and \ion{Ca}{II} K
with a narrower filter than the one used here are needed to better
understand what happens to the \ion{Ca}{II}~K fibrils above the quiet Sun.
 
Unlike H$\alpha$\ dynamic fibrils most of the \ion{Ca}{II} K fibrils do not
have well-defined endpoints even in the plage. This may be because
the fibrils (or perhaps more precisely the portion of the fibrils that
is visible in \ion{Ca}{II} K) are not in direct contact with the
transition region and the abrupt change in temperature associated with
it. This scenario has been proposed to explain why most of the nearly
horizontal H$\alpha$\ quiet Sun fibrils do not have clearly defined tops
\citep{DePontieu+others2007b}.

Many of the properties of the \ion{Ca}{II} K fibrils are similar to
fibrils observed in other spectral lines, such as H$\alpha$\ and
\ion{Ca}{II} 8542 \AA. The similarity of lengths and life times of H$\alpha$\ and \ion{Ca}{II} K fibrils was already noted by \cite{Marsh1976}.  
For example, also H$\alpha$\ fibrils
\citep{DePontieu+others2007b} and \ion{Ca}{II} 8542 \AA\ fibrils
\citep{Vecchio+others2009} are more dynamic in plage and active
network than in the quiet Sun. The dynamics of individual fibrils are
also similar: fibrils merge, split, fade away, etc. Since most of the
\ion{Ca}{II} K fibrils lack a well-defined top it is difficult to
measure velocities to see if they follow ballistic trajectories like many fibrils in 
H$\alpha$\ do \citep{DePontieu+others2007b}. The estimate of the \ion{Ca}{II} K
fibrils elongation speed, 20 km $\mathrm s^{-1}$, is in the velocity
range observed in H$\alpha$\ \citep{DePontieu+others2007b}. Large scale
movement along the fibrils away from magnetic concentrations is seen
in \ion{Ca}{II} 8542 \AA\ fibrils data as well
\citep{Cauzzi+others2008}.

The main difference between the \ion{Ca}{II} K fibrils and other
observations lies in the width of the fibrils. The \ion{Ca}{II} K fibrils are very
narrow, the average width being near the diffraction limit. One
of the more recent quotations on widths of dynamic fibrils seen in H$\alpha$\ is 310 km
(0.43 arcseconds, \citealt{DePontieu+others2007}). Spicules seen in
Hinode SOT \ion{Ca}{II} H data have widths of roughly 200 km, i.e., still larger than
the \ion{Ca}{II} K widths. The width of the SOT spicules, however, is
set by the spatial resolution of the instrument. The discrepancy
between H$\alpha$\ and \ion{Ca}{II} K may come from the fact that the \ion{Ca}{II} K
fibrils are a separate structure from H$\alpha$\ fibrils. Alternatively, the
horizontal extent of the emitting region might be smaller in \ion{Ca}{II} K than in H$\alpha$. Finally, the relatively broad filter available for the present observations may limit the apparent widths of the fibrils, since only little contribution is obtained from the chromospheric layers and only the bright spines of the fibrils are seen. 

The \ion{Ca}{II} K fibril width is related to the general magnetic
topology (e.g., brightness and length of fibrils) of the region but not
to the intensity of the fibrils. The wider fibrils in regions B
(plage) and D (weak plage) may be due to one or a combination of following possibilities: a) Noise: the faint fibrils
extending over the quiet Sun are more easily overshadowed by the
photospheric signal than the bright plage fibrils. b) The larger width
in the plage is a superposition effect c) The plage measurements were made
at the base of the fibrils where the fibrils (or at least the emitting
regions) are larger if one assumes that the fibrils (not
the flux tubes) become narrower with increasing distance from the
foot point. This scenario would take place for example if the fibrils get drowned in the background signal as discussed previously. Similar spatial dependence is seen in dynamic
H$\alpha$\ fibrils, where the median width in the plage is slightly higher
than at the edges of the plage (0.48 and 0.40 arcseconds,
respectively; \citealt{DePontieu+others2007}).
 
 The present observations show that the \ion{Ca}{II} K chromosphere is not as
different from H$\alpha$\ as first might appear but they also raise interesting questions; e.g. it is unclear how exactly the \ion{Ca}{II} K fibrils are related to
fibrils in other spectral lines. For example, are these the low-height
equivalent of H$\alpha$? Or are they the same structure? Are the
\ion{Ca}{II} K fibrils driven by magnetoacoustic shocks like a subset of the
H$\alpha$\ dynamic fibrils \citep{Hansteen+others2006}? This would explain
why a large portion of the \ion{Ca}{II} K fibrils tend to recur in the
same location whereas the other frequently proposed explanation, i.e.,
reconnection (e.g., \citealt{Uchida1969}), does not naturally account
for the recurrence. It is not unreasonable to conjecture that Ca II K
fibrils may also come in two varieties like the Hinode SOT \ion{Ca}{II} H
spicules \citep{DePontieu+others2007}, i.e., one consistent with
the shock scenario and the other more in line with reconnection. To
answer these questions more observations, preferably with a narrower filter or full spectra
and simultaneously with other spectral lines, are needed.

\section{Summary}

We have presented high resolution observations of the \ion{Ca}{II} K
line which show that very narrow fibrils are a prevailing feature in regions where the chromospheric signal is increased. Based on cotemporal continuum and nearly cotemporal
magnetic field observations it is clear that the fibril foot points
originate from photospheric magnetic concentrations. The fibrils share
many characteristics, e.g. lifetime and dynamics, with fibrils
observed in other chromospheric spectral lines. They play an important role in the
dynamics: in the plage they channel low frequency waves into the
chromosphere while in the more quiet regions the highly inclined
fibrils form a layered canopy that suppresses oscillations from
below.  

\acknowledgements{We wish to thank Han Uitenbroek for sharing his
  RH-code which was used to compute the \ion{Ca}{II} K contribution
  functions. The shock contribution function was computed of a
  snapshot from radiation hydrodynamic simulations by Carlsson \&
  Stein. We also wish to thank the referee for valuable comments. The Swedish 1-m Solar Telescope is operated on
  the island of La Palma by the Institute for Solar Physics of the
  Royal Swedish Academy of Sciences in the Spanish Observatorio del
  Roque de los Muchachos of the Instituto de Astrof$\acute{\rm i}$sica
  de Canarias. This work was partly supported by the WCU grant No. R31-10016 from the Korean Ministry of Education, Science and Technology.}


\end{document}